\documentclass[12pt]{article}
\usepackage{graphicx}
\usepackage{booktabs}
\usepackage{amsmath,amssymb,amsthm,graphicx,url}

\usepackage{times}
\usepackage{pdflscape}

\DeclareMathOperator{\agap}{\mathrm{Gap}_\tau}

\newcommand{\waste}{w}
\newcommand{\bp}{\boldsymbol{p}}

\newcommand{\incd}{I^D_i}
\newcommand{\incr}{I^R_i}
\newcommand{\wind}{W^D_i}
\newcommand{\winr}{W^R_i}

\newcommand{\mycapa}{Illustration of the angles $\theta_P$ and $\theta_Q$
    arising in the calculation of $\delta(\mathcal{E})\approx 0.54$
    for the $13$-district 2014 North Carolina congressional
    election. Districts have been sorted in increasing order of
    democratic vote share.}

\newcommand{\mycapb}{({\bf A}) In the $24$-district 1974 Texas congressional
    elections, the Democrats won approximately 68\% of the vote and
    87.5\% of the seats, yielding an efficiency gap of close to
    zero. The declination, however, is strongly negative as a
    consequence of marked change in slope of the distribution at
    $0.5$. 
    ({\bf B}) The vote distribution for the $18$-district 2012
    Pennsylvania congressional election. The mean district vote
    fraction for the Democrats is $0.504$. Also plotted is the
    regression line (slope $0.52$, intercept $0.24$; $r=0.875, p <
    0.001$) over all districts.
    The fact that it crosses the $0.5$ vote threshold approximately
    five districts earlier than the actual results do is consistent
    with $\delta_N = 4.8$.
    ({\bf C})
    Hypothetical election results for a district plan with 24
    districts whose democratic vote share ranges from $0.41$ in the
    southwest to $0.67$ in the northeast. The Democrats win $16$ seats
    with a mean vote share of $0.54$. 
    ({\bf D}) The vote distribution from the hypothetical
    election. The declination is near zero due to the regularity with
    which the democratic vote varies among districts. The efficiency
    gap is far from zero at $-0.17$ due to the fact that this election
    has a constant of proportionality close to one.}

\newcommand{\mycapc}{
  {\bf (A)} Change in declination relative to fraction of seats packed
  (California 2012 congressional election) or cracked (Arizona and
  Georgia 2012 congressional elections). Note that the elections
  referenced are used to provide an initial distribution to which
  packing/cracking is artificially applied.
    {\bf (B)} Plot of $N$ versus $|\tilde{\delta}|$ for 1\,472
    legislative and congressional elections. The regression line is
    $0.0007 N + 0.2011$ with an $r$-value of $0.1769$ and $p <
    0.0001$. 
}

\newcommand{\mycapd}{
Values of $\tilde{\delta}$ stratified by year for 461
    congressional elections {\bf (A)} and 606 state elections {\bf
      (B)} with at least eight districts (following~\cite{M-S}) and
    with each party winning at least one seat. Data for state
    elections after 2010 not collected. For each heat map, the number
    of included elections in a given year varies (for the state
    elections, states have gradually moved away from the multi-member
    districts (see Section~\ref{sec:mm}); for congressional elections there are
    sporadic instances in which one party wins all seats).
}

\newcommand{\mycape}{
Range of $\tilde{\delta}$ between 2012 and 2016 for
    congressional elections. Since Florida's congressional districts
    were changed for the 2016 election, that plan is denoted by FL2
    while the plan in place from 2012--2014 is denoted by FL1. States
    were omitted if one party won all seats every year of the cycle.
}

\newcommand{\mycapf}{
Comparison of the declination and (twice) the efficiency
    gap for $451$ state legislative elections from 1972 to
    2010. States with multi-member districts at any time since 1972
    are omitted. Declination (rather than $\tilde{\delta}$) is used
    even though we are comparing districts with different sizes in
    order to be consistent with the efficiency gap. The correlation is
    $r=0.76$ with $p < 0.001$. The three large black circles indicate the
    Wisconsin elections for 2012, 2014 and 2016.
}

\newcommand{\mycapg}{
{\bf (A)} Vote distribution for 1974 North Carolina
    congressional election with $\mathrm{Gap}_{0} = -0.1$ and
    $\mathrm{Gap}_{1} = -0.39$. {\bf (B)} Vote distribution for 2006
    Tennessee congressional election illustrating the mean-median
    difference of -0.13. {\bf (C)} Vote distrbution for the 2012
    Indiana congressional election showing a mean-median difference of
    0.01.
}

\newcommand{\mycaph}{
2016 congressional races with at least one win by each
    party. Listed in decreasing order of declination.
}

\newcommand{\mycapi}{
2008 state legislative races with at least one win by each
    party. Listed in decreasing order of declination.
}

\newcommand{\mycapj}{
Range of $\tilde{\delta}$ over each districting cycle for
    congressional elections: {\bf (A)} 1972--1980, {\bf (B)}
    1982--1990, {\bf (C)} 1992--2000, {\bf (D)} 2002--2010, {\bf (E)}
    2012--2016. Major redistrictings within each cycle are indicated
    by concatenation (e.g., TX1 refers to the district plan in effect
    at the beginning of the cycle and TX2 to the next plan). States
    were omitted if one party won all seats during the cycle. Subplot
    {\bf (E)} is identical to Fig.~\ref{fig:deltae-only} but is
    repeated here for comparison with earlier decades.
}

\newcommand{\mycapk}{
Range of $\tilde{\delta}$ over each districting cycle for
    state legislative elections: {\bf (A)} 1972--1980, {\bf (B)}
    1982--1990, {\bf (C)} 1992--2000, {\bf (D)} 2002--2010. Major
    redistrictings within each cycle are indicated by concatenation
    (e.g., OH1 refers to the district plan in effect at the beginning
    of the cycle and OH2 to the next plan). States were omitted if one
    party won all seats during the cycle (or if the plan had
    multi-member districts).
}

\title{Quantifying gerrymandering using the vote distribution} 

\author
{Gregory S. Warrington$^{1\ast}$\\
\\
\normalsize{$^{1}$Department of Mathematics \& Statistics, University of Vermont,}\\
\normalsize{16 Colchester Ave., Burlington, VT 05401, USA}\\
\\
\normalsize{$^\ast$To whom correspondence should be addressed; E-mail:  gswarrin@uvm.edu.}
}

\date{May 15, 2017}

\begin{document} 




\maketitle 

\begin{abstract}
To assess the presence of gerrymandering, one can consider the shapes
of districts or the distribution of votes. The \emph{efficiency gap},
which does the latter, plays a central role in a 2016 federal
court case on the constitutionality of Wisconsin's state legislative
district plan. 
Unfortunately, however, the efficiency gap reduces to proportional
representation, an expectation that is not a constitutional right.
We present a new measure of partisan asymmetry that does not rely on
the shapes of districts, is simple to compute, is provably related to
the ``packing and cracking'' integral to gerrymandering, and that
avoids the constitutionality issue presented by the efficiency gap. In
addition, we introduce a generalization of the efficiency gap that
also avoids the equivalency to proportional representation. We apply
the first function to US congressional and state legislative plans
from recent decades to identify candidate gerrymanders.
\end{abstract}


\section{Introduction}

To \emph{gerrymander} is to intentionally choose voting districts so
as to actively (dis)advantage one or more groups\footnote{There is no
  universally agreed upon definition. In particular, usage in the
  courts differs from colloquial usage.}. While gerrymandering can
focus on various types of groups, such as incumbents or those formed
by shared race, our focus in this article is the case in which the
disadvantaged group is a political party, that is, \emph{partisan
  gerrymandering}.

Approaches towards preventing gerrymanders include vesting the
responsibility of drawing districts in non-partisan commissions
(see~\cite{winburn2008realities}); drawing districts by putatively
neutral algorithms (such as the shortest splitline
method~\cite{splitline}); or moving away from the current
district-based single member plurality system (see~\cite{grofman} for
various alternatives). Since efforts along these lines have not yet
eradicated gerrymandering, there is a need for methods that can
accurately identify any gerrymanders that do occur.

Historically, gerrymanders have been identified by unusual district
boundaries~\cite{Griesbach}. In fact, it was the resemblance of a Massachusetts
senatorial district to a salamander that gives gerrymandering its name
(Elbridge Gerry signed off on the district in 1812 while governor of
Massachusetts). To this day, unexpected shapes of districts lead to
both zoomorphic comparisons (e.g., Pennsylvania's
6\textsuperscript{th} congressional ``dragon'' district)
and allegations of gerrymandering.

In response, a number of researchers have created compactness metrics
as a way to help identify gerrymandered districts
(see~\cite{compactness} for an overview); a recent, related
Markov-chain based approach is found in~\cite{pnas}. While such
approaches are important tools, the fact that partisan asymmetries
arise naturally from ``human geography''~\cite{chen} strongly suggests
gerrymanders can exist without contorted boundaries. By analogy, just
as one can be ill and yet not have a fever, so can one have a
gerrymander without violating compactness.

For more than four decades, vote-distribution asymmetries have been
analyzed via the ``seat-votes curve'' (see, for
example,~\cite{tufte,GK,KastellecGelmanChandler,nagle}), the computation
of which requires significant statistical assumptions.
In 2014, McGhee~\cite{McGhee} introduced an alternative, the
\emph{efficiency gap} (see also~\cite{M-S}). This function is easily
computed from its elegant definition without any assumptions at all
when all races are contested. A short derivation (see~\cite{McGhee} or
the derivation of equation~\eqref{eq:alphagap}) shows that a fair
district plan according to the efficiency gap is one in which the
seats won and the vote won are proportional with constant $2$: Winning
$53\%$ (that is, $50\% + 3\%$) of the vote will earn $56\%$ (that is,
$50\% + 2\cdot 3\%$) of the seats when the efficiency gap is zero.

As detailed in~\cite[III.C]{M-S}, several decades of court cases
predicated on partisan gerrymandering have, until recently, been
uniformly unsuccessful in the federal courts: On November 21, 2016, a
three-member circuit court panel ruled in \emph{Whitford
  v.\ Gill}~\cite{Wisconsin} that the districts drawn for the
Wisconsin lower house are an unconstitutional partisan
gerrymander. This is the first time such a determination has been made
regarding an alleged partisan gerrymander~\cite{nyt}. A significant
part of the panel's decision relies on the supporting evidence of
asymmetry afforded by the efficiency gap.  However, in his dissent in
\emph{Whitford v.\ Gill}, Chief Judge Griesbach objects to the
reliance on the efficiency gap, in large part on the grounds that
the US Supreme Court has ruled that \emph{there is no constitutional
  right to proportional representation}~\cite{Vieth}.  A natural
question is whether there are alternative functions that do not reduce
to proportional representation.

Below we do define a family of functions, indexed by nonnegative
numbers $\tau$, that specializes to (twice) the efficiency gap when
$\tau = 0$. As we'll describe in Section~\ref{sec:gap}, for $\tau > 0$
these functions do not reduce to proportionality. However, the focus
in this article will be on a new function for identifying
gerrymandering we term the \emph{declination}. It is a measure of
asymmetry in the vote distribution that relies only on the fraction of
seats each party wins in conjunction with the aggregate vote each
party uses to win those seats. (The declination is essentially an
angle associated to the vote distribution; its name is in analogy
with the angular difference between true north and magnetic north.)
The declination does not require the significant statistical
assumptions required to compute the seats-votes curve nor the
proportional-representation assumption embedded in the efficiency gap.

To be useful for identifying gerrymanders, our functions must measure
asymmetry that arises from gerrymandering rather than from other
sources. Asymmetries that are caused by partisan gerrymandering arise
from two primary techniques. The first is to \emph{pack} certain
districts with members of party $P$ so that party $P$ wins those
districts overwhelmingly, thereby wasting votes that could have been
helpful in winning districts elsewhere. A second technique, naturally
paired with the first, is to \emph{crack} the party-$P$ voters by
distributing them among districts so as to prevent them from having
sufficient power to win elections outside of the packed
districts. Overall, party $P$ wins a small number of overwhelming
victories while suffering a large number of narrow losses. In
Theorem~1 we rigorously prove that the declination
increases in response to packing and cracking.

In Section~\ref{sec:dec} we define the declination and two
variations. Analyses of elections in recent decades using the
declination are presented in Section~\ref{sec:elec}. The $\tau$-gap
family of functions that generalizes the efficiency gap is introduced
in Section~\ref{sec:gap}. In Section~\ref{sec:thm} we formally present
the theorem showing that both the declination and the $\tau$-gap
increase in response to packing and cracking. In
Section~\ref{sec:discuss} we discuss the relative merits of the
declination, the $\tau$-gap, the efficiency gap and the
mean-median. In Section~\ref{sec:mm} we outline the data sets used as
well as the statistical model used to impute votes in uncontested
elections. We conclude in Section~\ref{sec:conc}. Additional figures
and tables are included in the appendix.

\section{The declination}
\label{sec:dec}

Our definition of the declination stems from the observation that in a
randomly chosen (i.e., ungerrymandered) district plan, the $0.5$
threshold for the fraction of democratic votes in a district does not
play a distinguished role. Quirks of geography and sociology will
certainly shape the distribution in various ways that we will not
attempt to model. But we expect a phase shift across the $0.5$
threshold no more than we would expect one across $0.4$ or $0.56$.
Partisan gerrymandering, almost by definition, modifies a natural
distribution in a manner that treats the $0.5$ threshold as special.
Accordingly, one approach to recognizing gerrymanders is to contrast
the set of values below $0.5$ with the set of values above $0.5$.

\begin{figure}
  \centering
  \includegraphics[width=.8\linewidth]{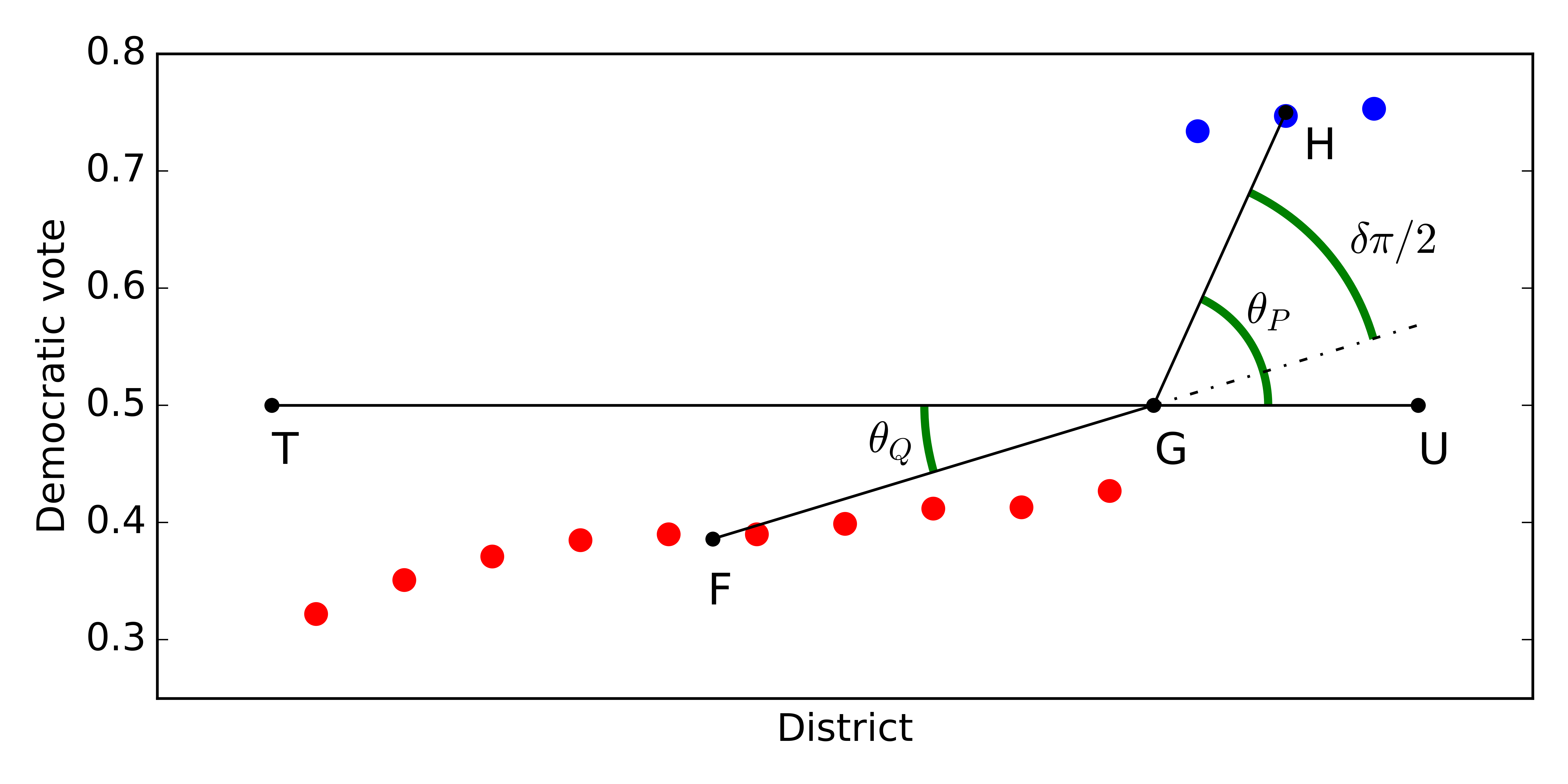}
  \caption{\mycapa}
  \label{fig:angle}
\end{figure}

In order to compare these two parts of the vote distribution, we
introduce some notation: Define an \emph{election with $N$
  districts} to be a triple $\mathcal{E} = (P,Q,\mathbf{p})$
consisting of political parties $P$ and $Q$ along with a vector
$\mathbf{p} = (p_1,p_2,\ldots,p_N)$ that records the fraction of the
two-party vote won by party $P$ in each of the $N$ districts. We
assume, without loss of generality, that our districts are ordered in
increasing order of party-$P$ vote share:
\begin{equation}\label{eq:p}
  0\leq p_1 \leq p_2 \leq \cdots \leq p_k \leq \frac{1}{2} < p_{k+1} \leq \cdots \leq p_{N} \leq 1.
\end{equation} 
Set $k'=N-k$ and
let $A = \{(i/N-1/2N,p_i):\, 1\leq i\leq k\}$ and $B =
\{(i/N-1/2N,p_{i}):\, k+1\leq i\leq N\}$.  Let $F=(k/2N,\bar{y})$ and
$H=(k/N+k'/2N,\bar{z})$ be the centers of mass of the points in $A$
and $B$, respectively (we therefore assume that $k,k'\geq 1$). Set $G
= (k/N,1/2)$, $T = (0,1/2)$ and $U=(1,1/2)$; see
Fig.~\ref{fig:angle}.  In a district plan in which neither side has
an inherent advantage, we would expect the point $G$ to lie on the
line $\overline{FH}$. Deviation of $G$ above $\overline{FH}$ is
indicative of an advantage for party $P$ while deviation below is
indicative of an advantage for party $Q$. This suggests, as a measure
of partisan asymmetry, the \emph{declination}, $\delta(\mathcal{E}) =
2(\theta_P-\theta_Q)/\pi$, where
\begin{equation}
  \theta_P = \angle HGU = \arctan\left(\frac{2\bar{z}-1}{k'/N}\right)\textrm{ and }
  \theta_Q = \angle FGT = \arctan\left(\frac{1-2\bar{y}}{k/N}\right).
\end{equation}
Dividing by $\pi/2$ converts from radians to fractions of $90$
degrees; possible values of the declination are between $-1$ and $1$.

To each party we are associating a line whose direction encodes the
average vote the party gets in the districts it wins along with the
total number of districts it wins. When a party uses lots of extra
votes to win few districts, the slope of this line is high. The
declination, up to the $\pi/2$-scaling, is the angle between the
lines associated to the two parties.

\begin{figure}
  \centering
  \includegraphics[width=.8\linewidth]{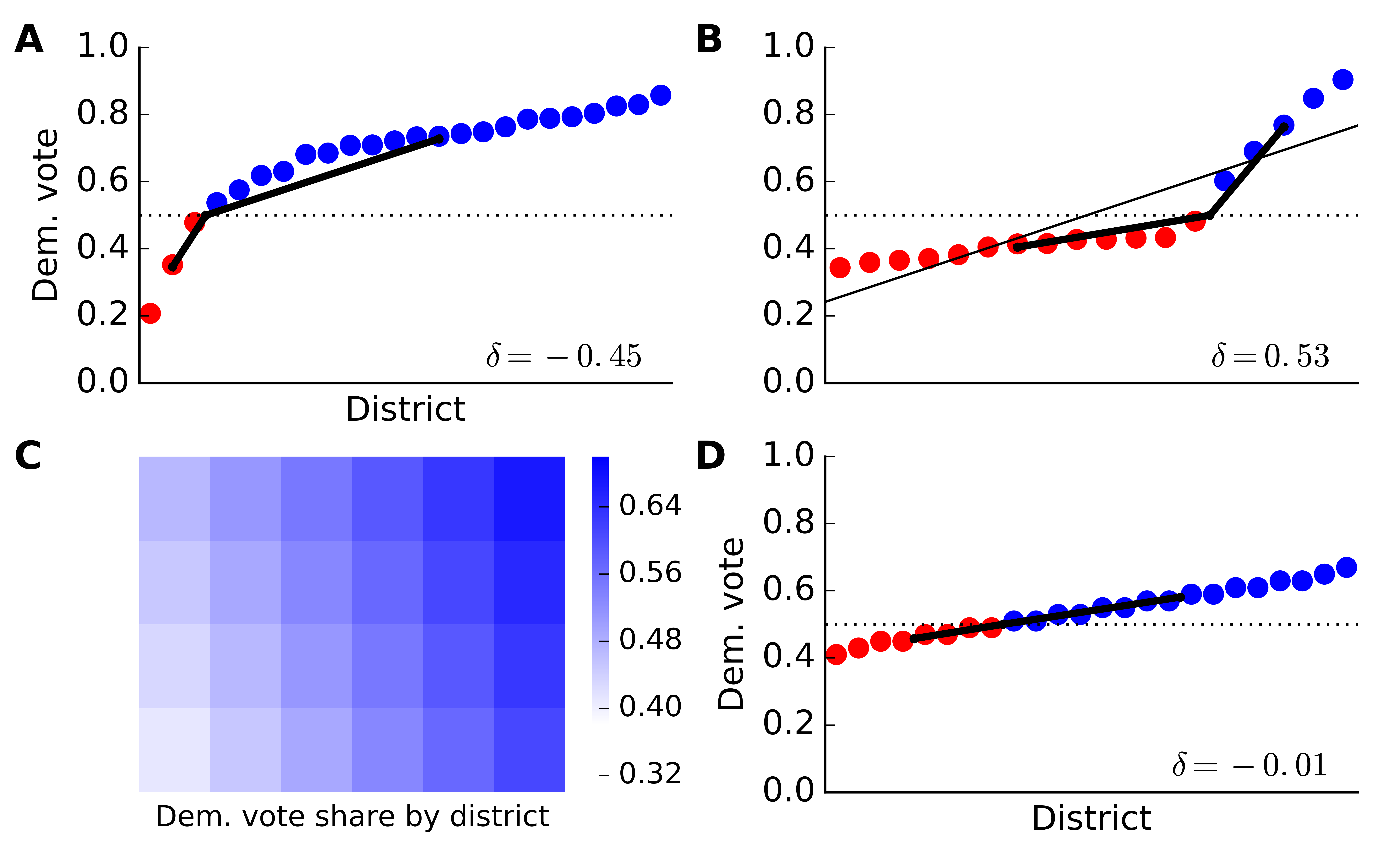}
  \caption{\mycapb}
  \label{fig:dec-ex}
\end{figure}

Fig.~\ref{fig:dec-ex} illustrates how the declination and the
efficiency gap evaluate different elections differently. In this
figure and the remainder of the article, we identify Party $P$ with
the Democrats and Party $Q$ with the Republicans. As a result,
positive values of the declination are favorable to Republicans.

\begin{figure}
  \centering
  \includegraphics[width=.8\linewidth]{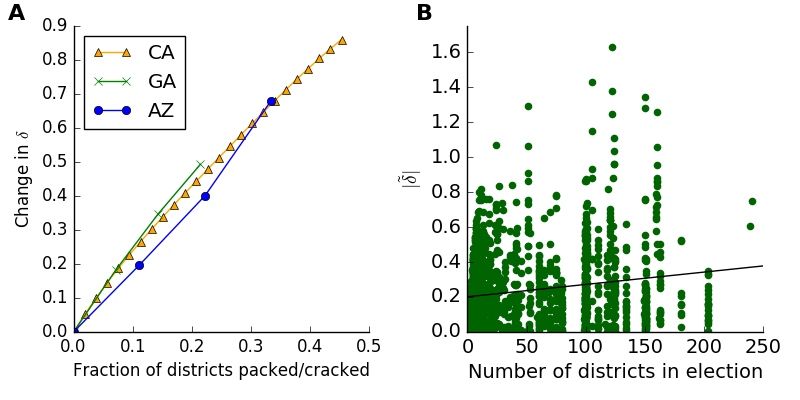}
  \caption{\mycapc}
  \label{fig:var}
\end{figure}

As illustrated in Figure~\ref{fig:var}.A, packing or cracking a
single district leads to a change of approximately $\pm 2/N$ in the
declination. Hence, as a measure of the number of seats affected by
the partisan advantage, we also consider $\delta_N(\mathcal{E})
=\delta(\mathcal{E})N/2$.

As a second variation, the value of $\tilde{\delta}(\mathcal{E}) =
\delta(\mathcal{E})\ln(N)/2$ has minimal correlation with $N$ (see
Fig.~\ref{fig:var}.B) and hence is useful when comparing the
declinations of elections with different numbers of districts.

\section{Analyses of elections}
\label{sec:elec}

\begin{figure}
  \centering
  \includegraphics[width=.8\linewidth]{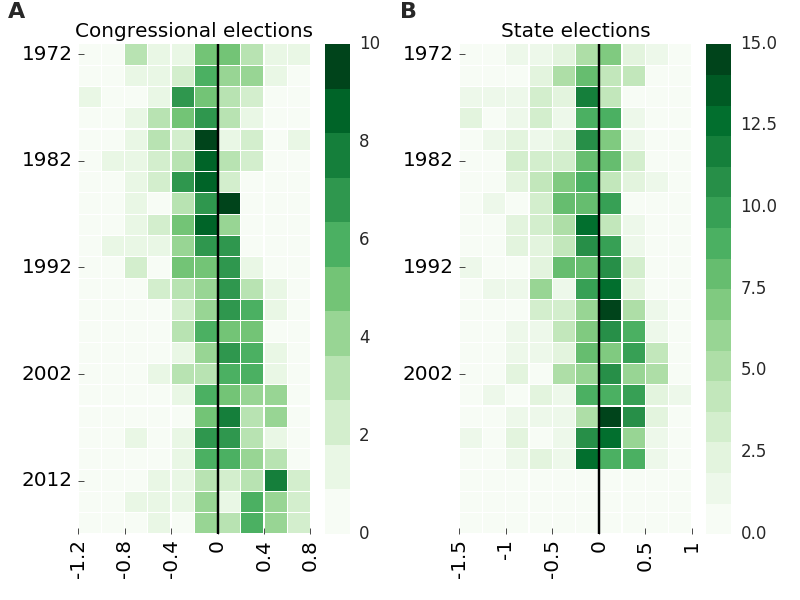}
  \caption{\mycapd}
  \label{fig:heatmaps}
\end{figure}

Fig.~\ref{fig:heatmaps} displays the distribution of $\tilde{\delta}$
values over time for state legislative (lower house) and congressional
races going back to 1972 (see Section~\ref{sec:mm} for a detailed
description of the data used).  We prove in Theorem~1 that the
declination increases (resp., decreases) as a result of gerrymandering
engendered by packing and cracking by party $Q$ (resp., party $P$).

\begin{table}[ht]
  \centering
  \caption{Values of the declination and its variants for the twenty
    2012 US Congressional elections in states with at least eight
    districts for which each party won at least one seat. Political
    control of the redistricting process taken from~\cite{cottrell}: R
    indicates republican control, D indicates democratic control, S
    indicates split control. Redistricting control in New Jersey and
    California was ostensibly non-partisan, but evidence suggests
    partisan control as indicated, see~\cite{cain,pierce} as noted
    in~\cite{cottrell}. Voting Rights Act (pre-clearance) states are
    marked by an asterisk.}
  \begin{tabular}{rrrcrl}\hline
    $\tilde{\delta}$ & $\delta$ & $\delta_N$ & State & No. Seats & Control \\\hline
 0.76 & 0.53 & 4.8 &  PA & 18 & R\\ 
 0.76 & 0.55 & 4.4 &  OH & 16 & R\\ 
 0.58 & 0.48 & 2.6 &  VA & 11 & R*\\ 
 0.58 & 0.45 & 2.9 &  NC & 13 & R\\ 
 0.56 & 0.43 & 3.0 &  MI & 14 & R\\ 
 0.49 & 0.44 & 2.0 &  IN & 9  & R\\ 
 0.46 & 0.28 & 3.8 &  FL & 27 & R\\ 
 0.44 & 0.24 & 4.4 &  TX & 36 & R*\\ 
 0.42 & 0.40 & 1.6 &  MO & 8  & R\\ 
 0.42 & 0.38 & 1.7 &  TN & 9  & R\\ 
 0.35 & 0.28 & 1.7 &  NJ & 12 & R\\ 
 0.32 & 0.31 & 1.2 &  WI & 8  & R\\ 
 0.28 & 0.21 & 1.5 &  GA & 14 & R\\ 
 0.16 & 0.09 & 1.3 &  NY & 27 & \quad S\\ 
 0.12 & 0.11 & 0.5 &  MN & 8  & \quad S\\ 
-0.08 &-0.07 &-0.3 &  WA & 10 & \quad S\\ 
-0.16 &-0.11 &-1.0 &  IL & 18 & \quad \quad D\\ 
-0.19 &-0.10 &-2.6 &  CA & 53 & \quad \quad D\\ 
-0.30 &-0.27 &-1.2 &  AZ & 9  & \quad S*\\ 
-0.55 &-0.53 &-2.1 &  MD & 8  & \quad \quad D\\\hline 
  \end{tabular}
  \label{tab:extreme}
\end{table}

There is no gold standard for assessing gerrymandering. As a result,
there is no direct way to validate the declination as a measure of
gerrymandering. However, we can still review the extent to which the
declination agrees with other measures of gerrymandering.
Table~\ref{tab:extreme} lists the values of $\tilde{\delta}$, $\delta$
and $\delta_N$ for congressional races in 2012. There is marked
consistency between the declination and who controlled the
redistricting process subsequent to the 2010 census. Also, while
disproportionality may not be sufficient evidence to denounce a
district plan as an unconstitutional gerrymander, a successful
gerrymander by definition results in one party winning fewer seats
than it would under a neutral plan. This is consistent with what
happened in the first four states in Table~\ref{tab:extreme}
(Pennsylvania, Ohio, Virginia and North Carolina): As noted
in~\cite{Wang}, the Democrats earned approximately half the statewide
vote in 2012 in each of these states while garnering no more than 31\%
of the seats in any of the four.

There is also strong agreement between the declination and compactness
metrics (we work here with $\tilde{\delta}$; $\delta$ and $\delta_N$ are
similarly concurrent). In~\cite{wp}, each 2012 congressional district
is scored by the Polsby-Popper compactness metric (scaled to run from
0 to 100; higher scores indicate less compact). The nine states with
at least eight districts and at least one district scoring greater
than 90 are North Carolina, Ohio, Pennsylvania, Virginia (the first
four listed in Table~\ref{tab:extreme}); California, Illinois and
Maryland (three of the last four in Table~\ref{tab:extreme}); as well
as Florida and Texas. 
Intriguingly, the declination identifies Indiana as a likely
gerrymander, in contrast to its evaluation in~\cite{wp} as a state
with very compact districts.

\begin{figure}
  \centering
  \includegraphics[width=.8\linewidth]{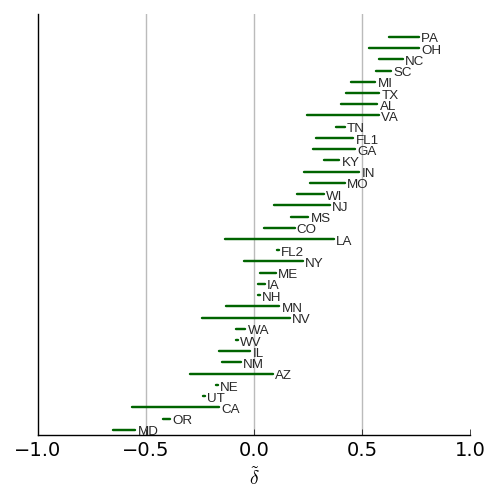}
  \caption{\mycape}
  \label{fig:deltae-only}
\end{figure}

The matter of determining a standard for what qualifies as an
unconstitutional partisan gerrymander is beyond the scope of this
article. (See~\cite{M-S} for a comprehensive treatment utilizing the
efficiency gap,~\cite{McGann} for the seats-votes curve
and~\cite{McDonaldBest,Wang} for the mean-median difference.) In
addition to the nontrivial task of addressing guidance from the
courts, two significant issues that must be addressed are the reality
that any measure of asymmetry in vote distributions will vary from
election to election (see Fig.~\ref{fig:deltae-only}) and that there may
be an inherent partisan asymmetry due to how voters are distributed
geographically. We briefly address each issue.

\begin{figure}
  \centering
  \includegraphics[width=.8\linewidth]{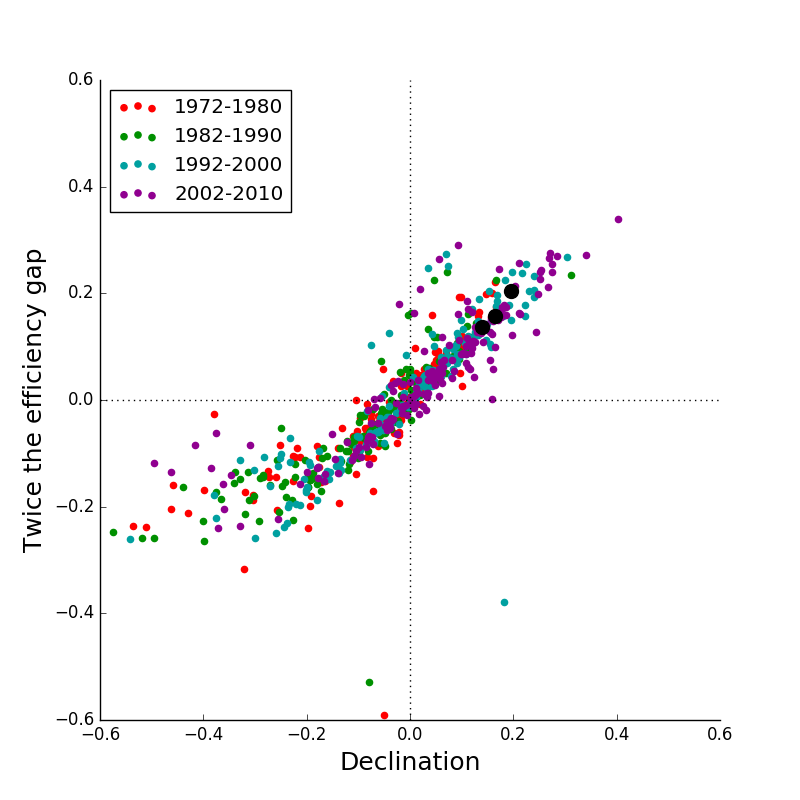}
  \caption{\mycapf}
  \label{fig:scatter-wi}
\end{figure}

\begin{figure}
  \centering
  \includegraphics[width=0.75\linewidth]{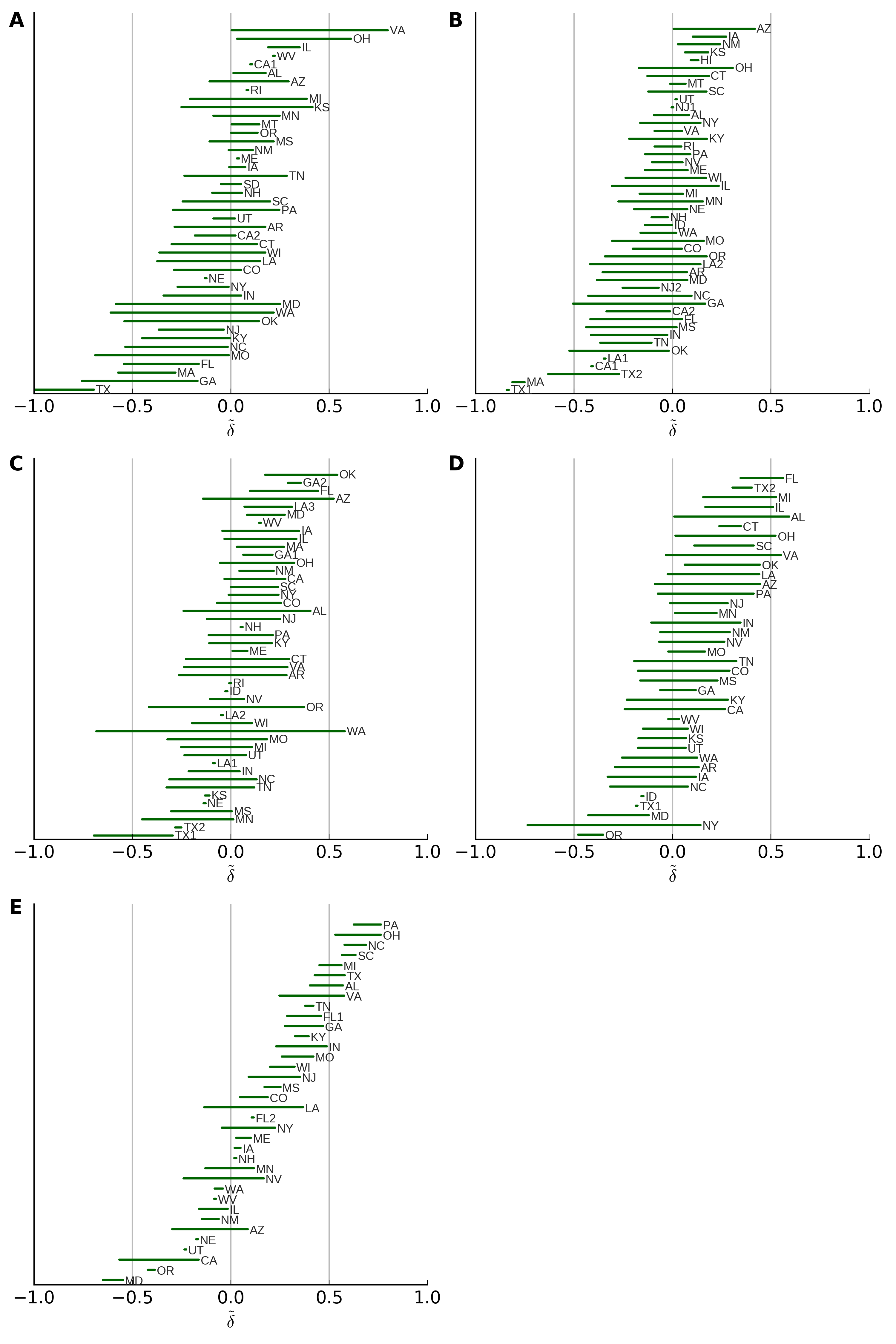}
  \caption{\mycapj}
  \label{fig:congint}
\end{figure}

\begin{figure}
  \centering
  \includegraphics[width=0.8\linewidth]{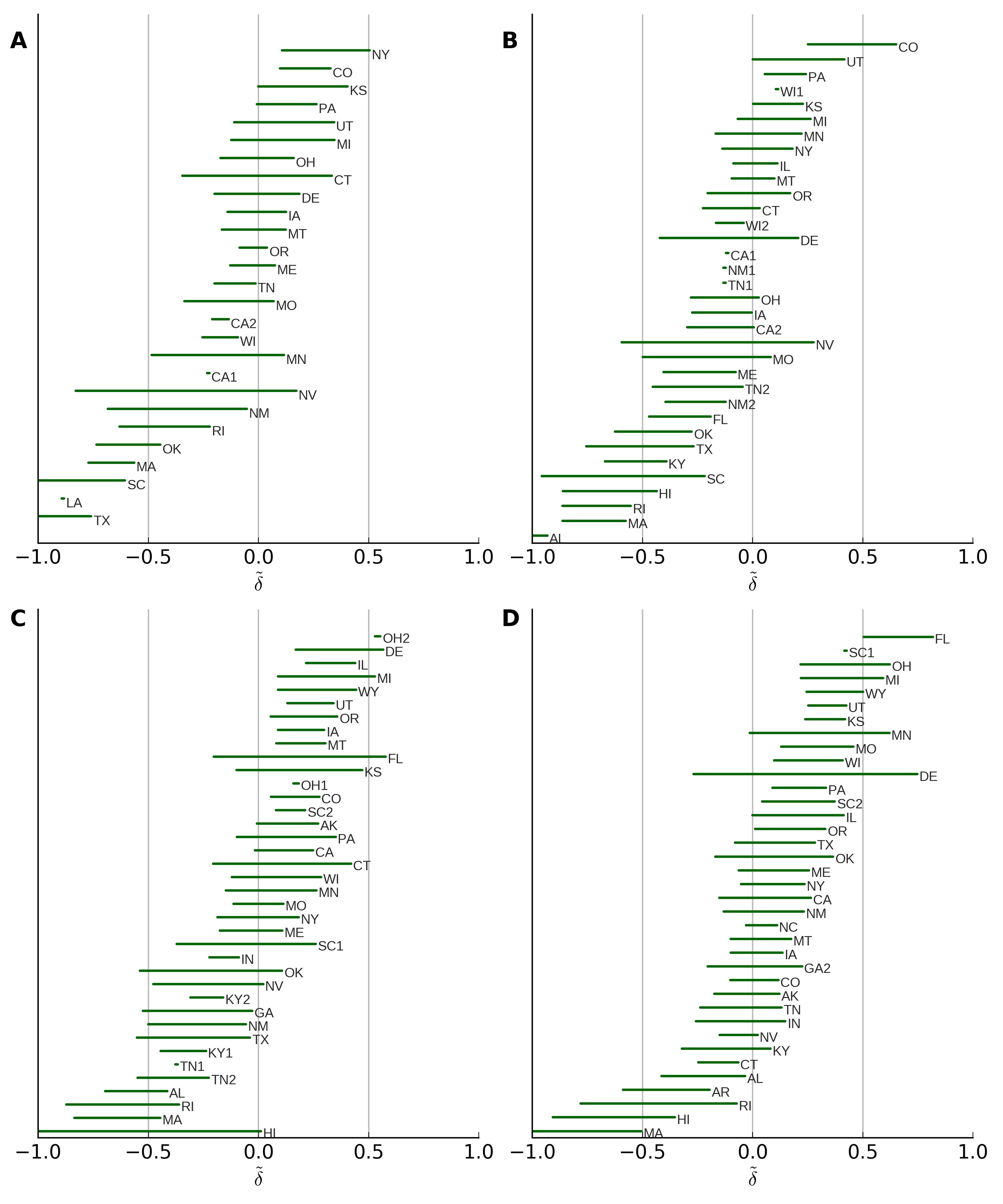}
  \caption{\mycapk}
  \label{fig:stateint}
\end{figure}

We consider the first problem by determining how large the declination
must be in absolute value before we are confident that it will not
equal zero for a different election in the same ten-year districting
cycle. Figs.~\ref{fig:congint} and~\ref{fig:stateint} illustrate the
total range of $\tilde{\delta}$ values for each state in each
redistricting cycle.  With respect to our historical data set of
1\,195 elections included in these figures, for over eighty percent of
the elections $\mathcal{E}$ with $|\tilde{\delta}(\mathcal{E})| >
0.47$, the declination (i.e., the partisan advantage) persists in sign
over the course of the entire ten-year redistricting cycle in which
$\mathcal{E}$ lies. With this confidence, therefore, we conclude from
Table~\ref{tab:extreme} (without examining data from 2014 or 2016)
that the partisan advantage for the congressional district plans in
Pennsylvania, Ohio, Virginia, North Carolina, Michigan, Indiana and
Maryland are likely to persist through 2020. We note that the 2012
Wisconsin legislative election, with $\tilde{\delta} = 0.48$, just
meets this threshold (also see Fig.~\ref{fig:scatter-wi}). For
reference, the most extreme values of the declination for
congressional races since 1972 are sorted in Table~\ref{tab:ext1972}.

The second major issue, inherent partisan advantages arising from how
voters are distributed geographically, is an important one. As
explored in~\cite{chen}, such realities may lead to one party
naturally garnering more than 55\% of the seats with only 50\% of the
vote. For example, if we take the republican advantage in Pennsylvania
to be 8\% (following~\cite{chen}), this seat advantage translates into
a value of $\tilde{\delta}$ of approximately $0.23$. Since $\tilde{\delta} = 0.76
> 0.47 + 0.23$ for the 2012 Pennsylvania congressional election, we
can still conclude with high confidence that the partisan advantage
for the Republicans \emph{beyond any geographic advantage} will
persist through 2020. Ultimately, the proper way to account for any
inherent geographic advantages requires answering the following: To
what extent is it constitutional for a district plan to exacerbate (or
mitigate) existing, natural advantages?

Additional summary tables and figures are included in the
appendix. Specifically,
Tables~\ref{tab:cong} and~\ref{tab:state} list the declination for all
1142 congressional elections and 646 state legislative elections in
the data set.
Figs.~\ref{fig:cong2016} and~\ref{fig:state2008} depict the actual
vote distributions (utilizing imputed vote values for uncontested
races (see Section~\ref{sec:mm}) along with the declination for the 2016
congressional elections and the 2008 state legislative elections,
respectively.

\section{A generalization of the efficiency gap}
\label{sec:gap}

In this section we introduce a variation on the efficiency gap
function. We first review the definition of the efficiency gap before
generalizing it in the next subsection.


McGhee defines a vote as \emph{wasted} if either 1) it
is a vote for the winning candidate in excess of the
one-more-than-50\% needed to win the election or 2) it is a vote for
the losing candidate. The motivation for this definition is that
packing and cracking produce more wasted votes in won and lost
districts, respectively. In light of this, the difference in the total
number of votes wasted by each party should reflect the extent to
which the packing and cracking treats the parties asymmetrically.

Given an election $\mathcal{E}=(P,Q,\boldsymbol{p})$, party $P$'s \emph{waste} in
district $i$, $w_P(i)$, is $p_i-1/2$ in those districts it wins
and $p_i$ in those districts it loses. Party $Q$'s waste is
$w_Q(i) = 1/2-w_P(i)$.

McGhee~\cite[Eq. (2)]{McGhee} (using different notation and terminology) made the
following definition: The \emph{efficiency gap} of $\mathcal{E}$ is 
\begin{equation}\label{eq:egap}
  \frac{\sum_{i=1}^N (w_P(i)-w_Q(i))}{N}.
\end{equation}

A positive (resp., negative) efficiency gap indicates that party $P$
(resp., $Q$) is wasting more votes. As such, ``fair'' district plans
should have efficiency gaps close to zero.


\subsection{The $\boldsymbol{\mathrm{Gap}_\tau}$ function}
The efficiency gap takes a binary view of votes: Either a vote
contributes to the first 50\% required for a candidate's win, and is
therefore not wasted at all, or it doesn't contribute, and is
therefore completely wasted. We now explore a more general paradigm in
which there is a continuum of waste that is shaped by a parameter
$\tau$.

As motivation for this generalization, consider the party-$Q$
gerrymanderers. They are not aiming to have the party-$P$ candidates
win districts by exactly one vote above 50\%. To do so would skirt
disaster. If their projections are even slightly too optimistic, then
the gerrymander will backfire spectacularly with each chosen
candidate losing by a razor-thin margin. It is less reckless for party
$Q$ to aim for comfortable margins in each district they aim to
win. Consequently, it is reasonable to consider the votes just above
50\% as only partially wasted while treating any votes near 100\% as
almost fully wasted.

For simplicity, we view the votes won by a given candidate as
ordered. For a winning candidate, the first half of the votes in the
district contribute directly to the win. The subsequent votes for the
winner, and all votes for the loser, are assigned waste values
determined by the parameter $\tau$. As will become apparent, setting
$\tau=0$ corresponds to weighting all wasted votes the same amount.

Given an $N$-district election $\mathcal{E}$, for each $i$, $1\leq
i\leq N$, set $a_i = 2p_i-1$. When $a_i$ is positive, party $P$ wins
district $i$ and $a_i$ is the fraction of the wasted votes that are
wasted by $P$. When $a_i$ is negative, party $Q$ wins district $i$ and
$-a_i$ measures the fraction of wasted votes wasted by party $Q$. 
For any nonnegative real
number $\tau$ and any $i > k$, we define the \emph{$\tau$-waste of
  party $P$ in district $i$} to be
\begin{equation}
  w_{P,\tau}(i) =
    \int_0^{a_i} x^{\tau}\,\mathrm{dx} = \frac{1}{\tau+1}a_i^{\tau+1}. 
\end{equation}
For $i\leq k$, similarly define $w_{Q,\tau}(i) =
(-a_i)^{\tau+1}/(\tau+1)$. The remaining values are
determined by requiring that the total waste in each district $i$,
$w_{P,\tau}(i) + w_{Q,\tau}(i)$, is constant and equal to
$\int_0^1 x^\tau\, \mathrm{dx} = 1/(\tau+1)$.
Set $w_{P,\tau} = \sum_{i=1}^N w_{P,\tau}(i)$
and $w_{Q,\tau} = \sum_{i=1}^N w_{Q,\tau}(i)$.
Finally, define the \emph{$\tau$-gap of election $\mathcal{E}$} to be
\begin{equation}\label{eq:main}
  \mathrm{Gap}_\tau(\mathcal{E}) = \frac{w_{P,\tau} - w_{Q,\tau}}{w_{P,\tau} + w_{Q,\tau}}.
\end{equation}
Note that for any election $\mathcal{E}$, $\mathrm{Gap}_{0}(\mathcal{E})$ is twice
the efficienct gap.

We now derive a formula for the $\tau$-gap that expresses it in terms
of the $a_i$. First we note that
\begin{align*}
  \waste_{P,\tau} &= \sum_{i=1}^k \waste_{P,\tau}(i) + \sum_{i=k+1}^N \waste_{P,\tau}(i)
  = \sum_{i=1}^k \left(\frac{1}{\tau+1} - \frac{(-a_i)^{\tau+1}}{\tau+1}\right) +
  \sum_{i=k+1}^N \frac{a_i^{\tau+1}}{\tau+1}\text{ and}\\
  \waste_{Q,\tau} &= \sum_{i=1}^k \waste_{Q,\tau}(i) + \sum_{i=k+1}^N \waste_{Q,\tau}(i)
  = \sum_{i=1}^k \frac{(-a_i)^{\tau+1}}{\tau+1} +
  \sum_{i=k+1}^N \left(\frac{1}{\tau+1} - \frac{a_i^{\tau+1}}{\tau+1}\right).
\end{align*}
Second, $\waste_{P,\tau} + \waste_{Q,\tau} = N/(\tau+1)$. Plugging
into equation~\eqref{eq:main}, we find that

\begin{align*}
  \agap(\mathcal{E}) &= \frac{\tau+1}{N}\left(-\sum_{i=1}^k2\frac{(-a_i)^{\tau+1}}{\tau+1} + \sum_{i=k+1}^N 2\frac{a_i^{\tau+1}}{\tau+1}
    + \frac{k-k'}{\tau+1}\right)\\
    &= 2\left(\frac{-\sum_{i=1}^k (-a_i)^{\tau+1} + \sum_{i=k+1}^N a_i^{\tau+1}}{N} + \frac{1}{2} - \frac{k'}{N}\right).
\end{align*}
It follows that
\begin{equation}\label{eq:alphagap}
  \mathrm{Gap}_\tau(\mathcal{E}) = 2\left[\frac{\sum_{i=1}^N \epsilon_i (\epsilon_i
      a_i)^{\tau+1}}{N} + \frac{1}{2} - \frac{k'}{N}\right],
\end{equation}
where $\epsilon_i = -1$ for $i\leq k$ and $\epsilon_i = 1$ for $i > k$.



We see that $\mathrm{Gap}_{0} = 0$ precisely when
\begin{equation}\label{eq:agapsimp}
\frac{\sum_{i=1}^N a_i}{N} = \frac{\sum_{i=1}^N 2p_i-1}{N} = 
\frac{k'}{N}-\frac{1}{2}.
\end{equation}
Let $\bar{p}$ denote the overall vote fraction won by party $P$.
Rewriting~(\ref{eq:agapsimp}), we find that the efficiency gap being
zero is equivalent to $2(\bar{p}-1/2)=k'/N-1/2$. We have recovered the
observation of~\cite{McGhee,M-S} that the efficiency gap is zero
exactly when the excess in fraction of seats won above one half is
twice the excess in vote fraction above one half.

For general $\tau$, however, the relationship between seats won and
votes earned in an election for which $\mathrm{Gap}_\tau$ is zero is more
complicated and depends on how the votes are distributed. When $\tau$
is a nonnegative even integer of the form $\tau = 2\ell$, we can
express the condition succinctly since then
\begin{equation}\label{eq:moment}
  \mathrm{Gap}_\tau(\mathcal{E}) = 2\left[M_{2\ell+1}(\boldsymbol{a}) + \frac{1}{2} - \frac{k'}{N}\right],
\end{equation}
where $M_{2\ell+1}(\boldsymbol{a})$ denotes the $(2\ell+1)$\textsuperscript{th} moment of the $a_i$.

In the limit as $\tau \rightarrow \infty$, $\mathrm{Gap}_\tau$ reduces (in the
general case of $0 < |a_i| < 1$ for all $i$) to $1-2k'/N$. It follows
that $\mathrm{Gap}_{\infty}$ is zero if and only if each party wins half of
the seats, irrespective of overall vote share.


There are at least two natural ways in which the $\tau$-gap could be
further generalized. The first is by allowing the waste to be most
pronounced close to the 50\% threshold. This is conveniently attained
by considering waste functions of the form $(1-a_i)^{-\tau}$ for
$\tau < 0$. This choice has several appealing characteristics such as
the fact that the winner and loser waste equal amounts more evenly
than the 75\%-25\% split for the efficiency gap. In addition, the
waste function is more sensitive to what is happening near the 50\%
threshold. Unfortunately, this latter property seems to reduce its utility
greatly as there is much more noise due to random fluctuations in vote
earned.

Second, while we have restricted our attention to power functions
determined by a parameter $\tau$, the concept of $\tau$-waste could
easily be extended to more sophisticated functions. One could also
allow different functions for the votes wasted by the winner and the
votes wasted by the loser.

\section{Theorem on packing and cracking} 
\label{sec:thm}

We make precise the link between packing and cracking and the
functions we introduce in the previous two sections. Define
\emph{$P$-cracking} to be the moving of party-$P$ votes from district
$k+1$ to districts $1,2,\ldots,k$ such that 1) the first $k$ districts
are still lost by party $P$ after the redistribution and 2) district
$k+1$ becomes a district that party $P$ loses. Similarly, define
\emph{$P$-packing} to be the moving of party-$P$ votes from district
$k+1$ to districts $k+2,k+3,\ldots,N$ such that district $k+1$ is now
lost by party $P$.  Using our new terminology, we state 
and prove the following

\noindent
{\bf Theorem 1.}\label{thm:main}
  Let $\mathcal{E} = (P,Q,\bp)$ be an election with $\bp$ in weakly
  increasing order. Let $\bar{y}$ be the average party-$P$ vote in the
  $k$ districts lost by party $P$ and $\bar{z}$ be the corresponding
  average in the $k'$ districts won by party $P$. Let $p_{k+1}'$ be
  the party-$P$ vote fraction remaining in district $k+1$ after
  $P$-cracking or $P$-packing and $\mathcal{E}'$ the resulting
  election. 
  \begin{enumerate}
    \item If $p_{k+1}' > \bar{y}$, then $\delta(\mathcal{E}') >
      \delta(\mathcal{E})$.
    \item If $p_{k+1}' > p_k$, then $\agap(\mathcal{E'}) >
      \agap(\mathcal{E})$.
  \end{enumerate}

\begin{proof}

  \emph{Proof of (1).} We refer the reader to Fig.~\ref{fig:angle} for
  the positions of points referenced. The $P$-packing or $P$-cracking
  of district $k+1$ will move point $G$ to the right by $1/N$
  units. Each of points $F$ and $H$ will move to the right by only
  $1/2N$ units. Since $p_{k+1}' > \bar{y}$, point $F$ will also move
  up. Since $\bp$ is arranged in increasing order, point $H$ will also
  move up. Together, these shifts will decrease $\theta_Q$ and
  increase $\theta_P$, thereby increasing $\delta$. By symmetry,
  $Q$-packing and $Q$-cracking will
  decrease $\delta(\mathcal{E})$. 

  \emph{Proof of (2).} We show only the details for $P$-cracking; the
  argument for $P$-packing is similar. Let $\bp' =
  (p_1',p_2',\ldots,p_N')$ be the vote distribution for
  $\mathcal{E}'$. Since we are $P$-cracking the
  $(k+1)$\textsuperscript{st} district, we know that $p_i'=p_i$ for $i >
  k+1$. Set $b_i = 2p_i'-1$ for $i\leq k+1$. Note in the below
  equations that all $a_i$ and $b_i$ for $i\leq k$ are \emph{negative}
  ($b_{k+1}$ is negative as well). Recall that
  \begin{align*}
    \agap(\mathcal{E}) &= 2\left[\frac{-\sum_{i=1}^k (-a_i)^{\tau+1} + a_{k+1}^{\tau+1} + \sum_{i=k+2}^N a_i^{\tau+1}}{N} + \frac{1}{2} - \frac{k'}{N}\right]\text{ and}\\
    \agap(\mathcal{E}') &= 2\left[\frac{-\sum_{i=1}^k (-b_i)^{\tau+1} - (-b_{k+1})^{\tau+1} + \sum_{i=k+2}^N a_i^{\tau+1}}{N} + \frac{1}{2} - \frac{k'-1}{N}\right].
  \end{align*}
  We wish to show that the difference
  \begin{equation*}
    \agap(\mathcal{E}')-\agap(\mathcal{E}) = 
\frac{\sum_{i=1}^k(-(-b_i)^{\tau+1}+(-a_i)^{\tau+1}) - (-b_{k+1})^{\tau+1} - a_{k+1}^{\tau+1}}{N} + \frac{1}{N}
  \end{equation*}
  is positive. Multiplying by $N/(\tau+1)$, we see that this is equivalent to showing that
  \begin{equation*}
    \frac{\sum_{i=1}^k ((-a_i)^{\tau+1}-(-b_i)^{\tau+1}) + 1}{\tau+1} \geq \frac{(-b_{k+1})^{\tau+1} + a_{k+1}^{\tau+1}}{\tau+1}.
  \end{equation*}
  We make the following observations. First, since all votes being
  moved are going from district $k+1$ to a district $i$ with $i\leq
  k$, we must have $\sum_{i=1}^k ((-a_i)+b_i) =
  a_{k+1}-b_{k+1}$. Second, since $a_{k+1}$ is positive $\sum_{i=1}^k
  ((-a_i)+b_i) \geq -b_{k+1}$.  So choose constants $0 = c_0 \leq c_1
  \leq c_2 \leq \cdots \leq c_k = -b_{k+1}$ such that $(-a_i)-(-b_i)
  \geq c_i-c_{i-1}$ for $1\leq i\leq k$. Straightforward calculus shows that,
  since $x^{\tau}$ is an increasing function and $-b_{k+1} \leq -a_i$
  for all $i\leq k$ by hypothesis, that
  $((-a_i)^{\tau+1}-(-b_i)^{\tau+1}) \geq
  c_i^{\tau+1}-c_{i-1}^{\tau+1}$ for all $1\leq i\leq k$. It follows
  that 
  \begin{equation*}
    \sum_{i=1}^k \int_{-b_i}^{-a_i} x^\tau\,\mathrm{dx} =
    \sum_{i=1}^k \frac{((-a_i)^{\tau+1}-(-b_i)^{\tau+1})}{\tau+1} \geq \frac{(-b_{k+1})^{\tau+1}}{\tau+1}.
  \end{equation*}
  As $1/(\tau+1) \geq a_{k+1}^{\tau+1}/(\tau+1)$, the result follows.
\end{proof}

\section{Strengths and weaknesses of different functions}
\label{sec:discuss}

Our situation in identifying gerrymanders is analogous to that of
assessing intelligence. Emotional intelligence and logical
intelligence are best measured by different types of questions; no
single question type can capture the spectrum of ways in which
intelligence can manifest itself. Every test for gerrymandering will
have its strengths and weaknesses. In this section we attempt to
illustrate, through concrete examples, ways in which each of the
efficiency gap, the $\tau$-gap, the mean-median difference and the
declination can give misleading answers in their attempts to identify
gerrymanders.

The oldest tool for this purpose, the seats-votes curve, has been
around for over 40 years~\cite{tufte}. Notwithstanding its theoretical
interest and applications in such areas as predictions (for example,
~\cite{KastellecGelmanChandler}), as it has not led to a manageable
judicial standard for gerrymandering in this span, we will not address
it directly in the remainder of this article.

\subsection{The efficiency gap}
\begin{figure}
  \centering
  \includegraphics[width=0.8\linewidth]{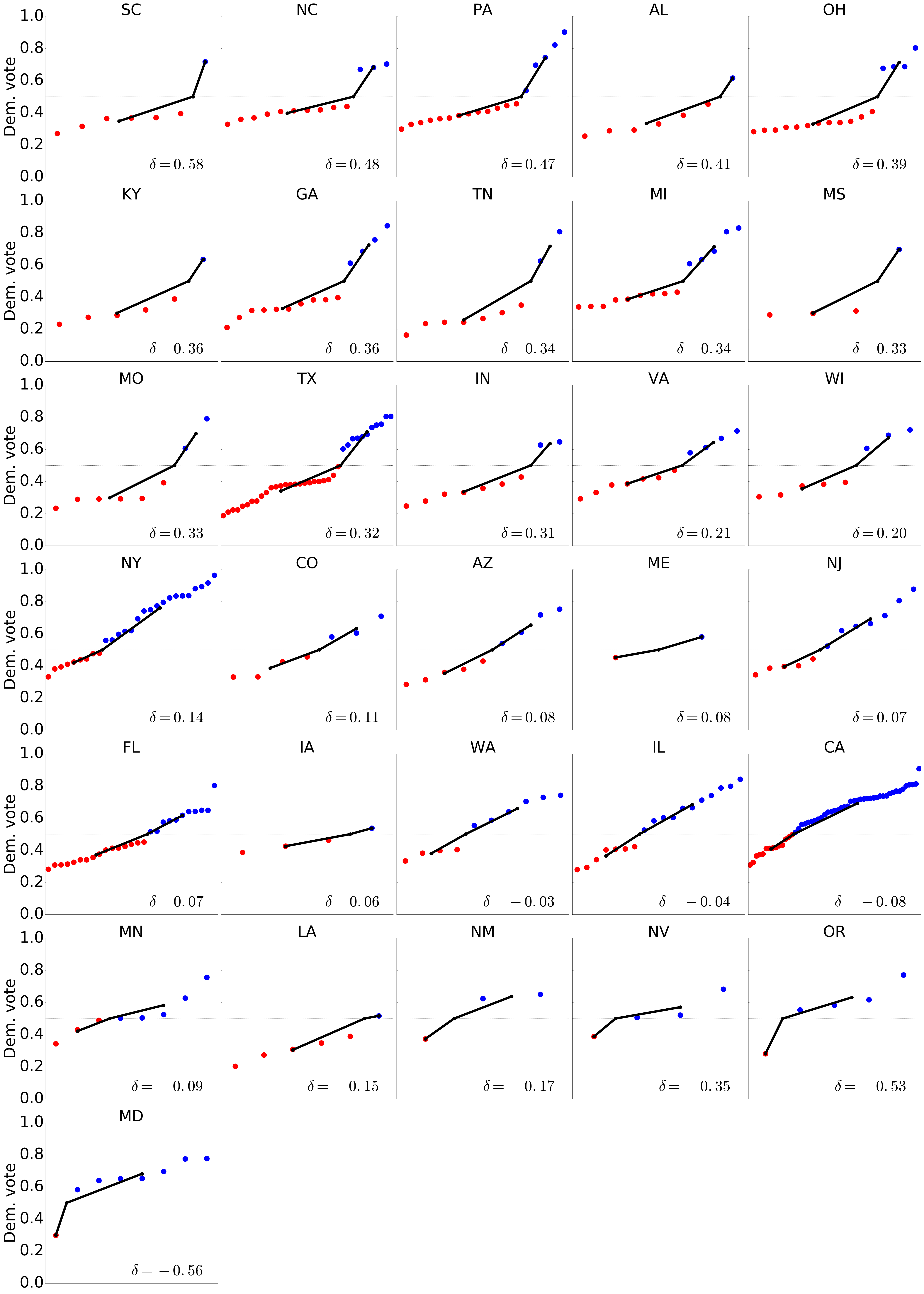}
  \caption{\mycaph}
  \label{fig:cong2016}
\end{figure}

The efficiency gap, through its role in \emph{Whitford v. Gill}, has
already proven its utility. And while its simplicity is an asset, it
also has several drawbacks. First, by requiring that there be
proportional representation for an election to be considered fair, it
conflicts with constitutional law~\cite{Vieth}. Second, it is simple
to construct hypothetical examples for which a natural district plan
leads to a constant of proportionality different from $2$ (see, for
example, Figs.~\ref{fig:dec-ex}.C,D). While the proportionality
asserted by the efficiency gap may hold in some overall sense, for the
instances for which it doesn't we are left in the difficult position
of determining whether any deviation from this average law is due to
gerrymandering or natural deviation. And, as illustrated by our
definition of the $\tau$-gap function, the proportionality with a
constant of $2$ depends on a subjective decision on how to weight
various wasted votes. Third, there are historical examples in which
the proportionality holds, but the shape of the vote distribution
suggests significant partisan asymmetry (see, in particular, the 1974
Texas congressional election shown in Fig.~\ref{fig:dec-ex}.A as well
as the 2012--2016 Tennessee congressional elections, the last of
which is displayed in Fig.~\ref{fig:cong2016}).

%

\subsection{The $\tau$-gap}
The $\tau$-gap family of functions is appealing due to its close
connection to the efficiency gap and the fact that it does not reduce
to proportionality. Additionally, when $\tau=2$, the $\tau$-gap is
closely related to skewness of the $a_i$-distribution
(see~\eqref{eq:moment}), a standard measure of symmetry in statistical
distributions. We also note that when $\tau=2/5$,
the correlation between the declination and the $\tau$-gap is very
high ($r^2=0.870, p < 0.001$ among $461$ congressional
elections). Unfortunately, the $\tau$-gap family has its own
drawbacks.

\begin{figure}
  \centering
  \includegraphics[width=.8\linewidth]{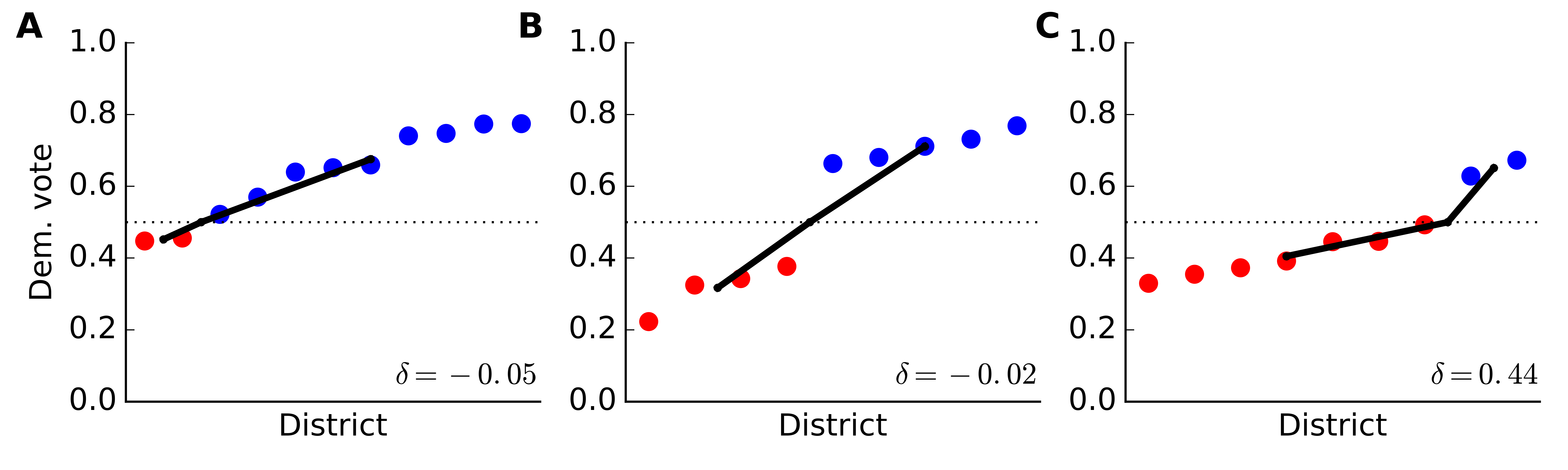}
  \caption{\mycapg}
  \label{fig:discuss}
\end{figure}

First, there is no reason to think that one value of $\tau$ will prove
superior to other values in all cases. For example, consider the 1974
Texas and North Carolina congressional elections (see
Fig.~\ref{fig:dec-ex}.A and Fig.~\ref{fig:discuss}.A, respectively).
Proportionality is essentially met in both cases as reflected by the
values of $\mathrm{Gap}_{0}$ being close to zero; each election is
fair according to the efficiency gap. In contrast, $\mathrm{Gap}_{1}$
is approximately $-0.38$ for each, indicating strong asymmetry.  While
we have no absolute way of determining whether one or both is a
gerrymander, evidence suggests that $\mathrm{Gap}_0$ makes the correct
assessment for the North Carolina election (little or no
gerrymandering) while $\mathrm{Gap}_1$ makes the correct assessment
for the Texas election (gerrymandering)\footnote{As described
  in~\cite[pg. 35]{ncredistrict}, there was not significant partisan
  rancor during the 1971 redistricting and the Democrats received
  approximately 63\% of the congressional vote while winning 82\% (9
  of 11) congressional seats. Of course, these facts have little
  bearing on whether there may have been \emph{racial}
  gerrymandering. For the Texas election, one author~\cite{texas}
  suggests that for the 1965 redistricting, protecting incumbents was
  a higher priority than advantaging the Democrats as a party. The
  constructions of a large number of safe seats for incumbents would
  certainly be consistent with the distribution seen.}. We note that
the declination yields what we believe to be the correct answer in
both cases.

Second, even if there were one value of $\tau$ that was superior to
all others, it is unclear what theoretical, rather empirical,
justification could be used to support it. Third, as shown
by~\eqref{eq:alphagap}, the $\tau$-gap is essentially an
interpolation, albeit a complicated one, between the proportionality
with constant $2$, when $\tau=0$, and proportionality with constant
$0$, when $\tau=\infty$. (A fair election for the latter is one in
which each party gets half the seats, regardless of the fraction of
votes won by each side.)

\subsection{The mean-median difference}
The difference between the mean and the median of the democratic vote
fraction among all districts has been suggested as yet another way of
measuring partisan asymmetry (see, for example, Wang~\cite{Wang} and
\cite{krasno,McDonaldBest}).  A large difference between these values
can indeed accompany an asymmetry in the vote distribution. For
example, in the 2012 Pennsylvania congressional election (see
Fig.~\ref{fig:dec-ex}.B), the mean is 0.50 while the median is 0.43, a
difference of 0.07. However, the mean-median difference is very
sensitive to the vote in a single district (practically speaking, this
is less of an issue in the case of many districts). For example, in
the 2006 Tennessee congressional election shown in
Fig.~\ref{fig:discuss}.B, the difference is -0.13. Had the democratic
support been more unevenly distributed in the districts the Democrats
won, the difference would have been much smaller in absolute
value. The declination, by relying on averages across districts rather
than values within single districts is more robust in this respect.

Additionally, it is quite possible for packing and cracking to occur
without having any effect on the mean-median difference. For example,
consider the 2012 Indiana congressional election shown in
Fig.~\ref{fig:discuss}.C. According to the mean-median difference,
this is a fair election. However it is easy to see how less dominant
wins in two districts could have translated into one or even two more
seats for the Democrats: such changes need not affect the median.

It is also worth noting that the mean-median difference doesn't keep
track of the number of seats won by each party. This is probably the
main reason it is slightly more stable from election to election than
the efficiency gap or the declination. On the other hand, this
independence makes it harder to interpret the extent to which a
particular value of the difference indicates actual impact on the
electoral results.

\subsection{The declination}
\begin{figure}
  \centering
  \includegraphics[width=0.8\linewidth]{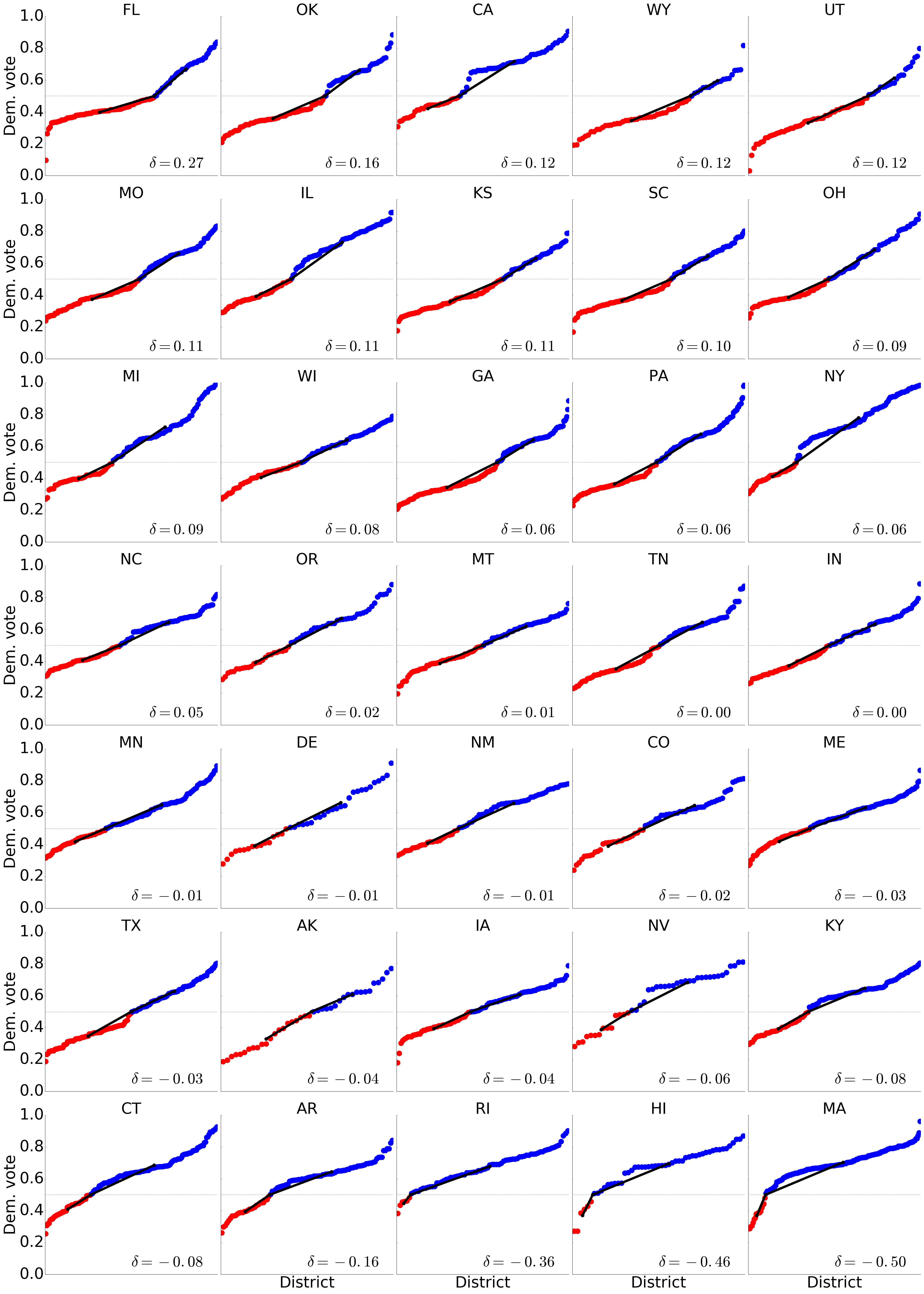}
  \caption{\mycapi}
  \label{fig:state2008}
\end{figure}

The declination's primary weaknesses appear to occur when one party
wins almost all of the seats. Fortunately, while gerrymandering may
occur in such instances, these are typically not the cases of greatest
interest.  In most elections with a large number of seats, the vote
fractions when sorted in increasing order increase essentially
linearly except at the extremes. When one party dominates, this can
lead to a declination that is large in absolute value even though the
vote distribution as a whole is remarkably symmetric about the median
democratic votes share. This is the case, for example in the 2008
Rhode Island state legislative election (shown in
Fig.~\ref{fig:state2008}), in which the Democrats win two thirds of the
vote. On the other hand, when there are few seats overall, it is not
uncommon for one party to win all but one seat. In such a case, the
declination is very sensitive to the exact vote fraction found in that
one special district.





\section{Data collection and Statistical methods}
\label{sec:mm}

The US state legislature election data up through 2010 comes
from~\cite{carsey}. We only included data on the lower house of the
state legislature when two houses exist. The US congressional data
through 2014 was provided by~\cite{jacobson}. Data on
Wisconsin legislative elections from 2012 to 2016 were take from
Ballotpedia~\cite{ballot}.  Data for 2016 congressional races were
taken from Wikipedia~\cite{wiki:2016} as they were not yet available
from the Federal Election Commission.

The election data was analyzed using the python-based
SageMath~\cite{sage}.
Python packages employed were pyStan~\cite{pystan} for implementing a
multilevel model to impute votes in uncontested races;
Matplotlib~\cite{matplotlib} and Seaborn~\cite{seaborn} for plotting
and visualization; and SciPy~\cite{scipy} for statistical methods.

For the state legislative elections, we restricted the analysis to elections that
had no multi-member districts; vote fractions for uncontested races
were imputed (i.e., we estimated what the values would have been had
the races been contested).

Any election (year and state) in which there is at least one
multi-member district (i.e., multiple winners in a single district)
was excluded from our analysis. We ignored any third-party candidates
and assumed there to be at most one Democrat and at most one
Republican in each race. All vote percentages assigned to a candidate
are therefore percentages of the two-party vote. We assume all
districts have equal population and we do not take into account voter
turnout in any way. Unopposed candidates are allowed. The declination
is mildly sensitive to the vote distribution in each of these
uncontested races (see below). Rather than assign 100\% of the vote to
the unopposed candidate, which is invariably not reflective of what
would have happened had their been a two-candidate race, we imputed
the vote fraction for such races.

Imputation of votes was done using a multilevel model of the form
\[
  y_i = \frac{1}{2} + \sigma_{j[i]} + \phi_{k[j[i]]} + \gamma_{\ell[i]} + \\
  \beta_1 \wind + \beta_2 \winr + \beta_3 \incd + \beta_4\incr + \epsilon_i.
\]
Here, $y_i$ is the fraction of the vote garnered by the democratic
candidate in district $i$; $\sigma_{j[i]}$ is a random state effect;
$\phi_{k[j[i]]}$ is a random district effect; $\gamma_{\ell[i]}$ is a
random year effect; $\wind$ and $\winr$ are indicator variables for
whether the democratic or republican candidate won, respectively;
$\incd$ and $\incr$ are indicator variables for whether or not the
democratic and republican candidates are incumbents, respectively;
$\beta_1$, $\beta_2$, $\beta_3$ and $\beta_4$ are the corresponding
random effects; and $\epsilon_i$ represents error due to individual
characteristics of race $i$. The model was run separately for each
redistricting cycle (typically of ten years), once for state
legislatures and once for the US House of Representatives. As there
are a number of cases in which states underwent major redistricting
mid-decade, certain cycles were split. For example, Texas had one
district plan for its congressional races during the years 1992--1996
and another for 1998--2000. These are represented in the
Figure~ref{fig:congint}.C by TX1 and TX2, respectively.

There were 646 state elections in our data with a total of 68\,955
races of which 25\,371 of which were uncontested. There were 1\,142
congressional elections with a total of 9\,995 races of which 1\,409
were uncontested.

When at least one race in a given district had been contested in a
given cycle, imputation is straightforward using the values for the
random and fixed effects provided by the model fit. We cross-validated
the estimates by removing individual races, refitting and comparing
the estimated to true values. We found a root mean square error of
approximately $0.05$ among $100$ randomly chosen contested races. We
note for comparison that if one uniformly assigns 65\% of the vote to
the winner, the error is approximately $0.09$.

Of the 26\,780 imputed values, there were 114 instances in which the
data indicated a democratic winner, but the imputed value was less
than or equal to $0.50$. These imputed values were replaced with
$0.505$. Similarly, there were 40 instances in which a democratic
loser was indicated, but the imputed value was greater than
$0.50$. These imputed values were replaced with $0.495$.





There a number of instances in which a district was not
contested at any time during a given reapportionment cycle. There were
11\,770 state legislative district-cycle pairs. Of these, 1\,281 were
uncontested (i.e., the race in that district was not contested in any
year of the cycle). The corresponding numbers for congressional
elections are 2\,030 and 41. For the vast majority of these cases, one
of the two parties held the seat for the entire cycle. For these
cases, we drew from the distribution of district effects stemming from
districts that were consistently won by the same party throughout the
cycle (with at least one contested race). There were a total of 19
state district-cycle pairs that were uncontested but held by both
parties at some point during the cycle. The district effect in these
cases was drawn at random from all district effects. This outcome did
not occur for the congressional races.


We also performed a sensitivity analysis with respect to the imputed
values by introducing a systemic bias of plus $3\%$ to the democratic
imputed votes for congressional elections. We then performed a linear
regression of change in declination with respect to the fraction of
races that had to be imputed. Elections in which one party won all of
the seats were omitted. The slope and $r^2$-value for the regression
line was $(0.09,0.76)$. The corresponding regression line for state
elections was similar, but with a higher $r^2$-value.

For congressional elections with at least eight seats, 53\% had at
most 10\% of the races uncontested while 90\% had at most 40\% of the
seats uncontested. For state elections, only 31\% had less than 10\%
of the seats uncontested while 92\% had less than 65\% of the races
uncontested.

\section{Conclusion}
\label{sec:conc}

Among the tests discussed in the previous section, we believe that the
declination, on the whole, possesses the most desirable combination of
characteristics. First, it does not reduce to proportionality, and
hence is not at odds with constitutional law. Second, it is readily
visualized. Third, the declination is directly computed using
fundamental aspects of the election, namely the number of seats each
party wins along with the average vote fraction each earns in those
wins. While the efficiency gap is perhaps even simpler, the general
$\tau$-gap is certainly much harder to visualize. Fourth, the
declination is relatively robust with respect to the vote fraction in
any individual district, unlike the mean-median difference. The
declination also must change in the presence of packing and cracking,
a feature that does not always hold for the mean-median difference.

The declination and compactness metrics are complementary, each
captures particular characteristics of gerrymandering. These measures
do not capture everything about an election and one must be careful to
not ascribe more importance to a single number than is warranted
(see~\cite{cathy}).  Nonetheless, they have utility through their
ability to provide a consistent way to compare elections in different
states and years. Assuming the courts ultimately accept one or more
standards for ascertaining whether a district plan amounts to
unconstitutional gerrymandering, it will be helpful if there is enough
flexibility provided so that as our tools and understanding evolve, so
do our classification standards.


\section{Acknowledgments}
This work was partially supported by a grant from the Simons
Foundation (\#429570). The data reported in this paper are archived at
the following databases (TBD). The author is especially indebted to
Gary C. Jacobson for sharing his data on US Congressional
elections. The author also gratefully acknowledges extensive
discussions with Jeff Buzas and Jill Warrington as well as helpful
input from Jim Bagrow, Sara Billey, Chris Danforth, Jeff Dinitz, Mark
Moyer, John Schmitt, Mike Schneider, Dan Velleman and \{Ann, Bob,
Jeff\} Warrington.


\clearpage

\section{Appendix}


\newpage

\newpage

\newpage

\newpage

\newpage

\begin{table}[ht]
  \centering
  \caption{Most extreme values of $\tilde{\delta}$ for congressional elections since 1972.}
  \begin{tabular}{@{}ccrrrrcccrrrr@{}} \hline
 \multicolumn{6}{c}{Most positive $\tilde{\delta}$} & & 
 \multicolumn{6}{c}{Most negative $\tilde{\delta}$}\\
 \cmidrule{1-6} \cmidrule{8-13}
Year & State & Seats & $\tilde{\delta}$ & $\delta$ & $\delta_N$ & &
Year & State & Seats & $\tilde{\delta}$ & $\delta$ & $\delta_N$\\\hline
1980 & VA & 10 &  0.80 &  0.69 & 3.5 & & 1976 & TX & 24 & -1.07 & -0.67 & -8.1\\
2012 & PA & 18 &  0.76 &  0.53 & 4.7 & & 1982 & TX & 27 & -0.83 & -0.51 & -6.8\\
2012 & OH & 16 &  0.76 &  0.55 & 4.4 & & 1990 & MA & 11 & -0.82 & -0.68 & -3.7\\
2014 & NC & 13 &  0.69 &  0.54 & 3.5 & & 1988 & MA & 11 & -0.79 & -0.66 & -3.6\\
2016 & PA & 18 &  0.67 &  0.47 & 4.2 & & 1984 & MA & 11 & -0.77 & -0.65 & -3.5\\
2012 & SC & 7 &  0.63 &  0.65 & 2.3 & & 1986 & MA & 11 & -0.77 & -0.64 & -3.5\\
2014 & PA & 18 &  0.62 &  0.43 & 3.9 & & 1972 & GA & 10 & -0.76 & -0.66 & -3.3\\
2016 & NC & 13 &  0.61 &  0.48 & 3.1 & & 1982 & MA & 11 & -0.75 & -0.63 & -3.5\\
1972 & OH & 23 &  0.61 &  0.39 & 4.5 & & 2008 & NY & 29 & -0.74 & -0.44 & -6.3\\
2010 & AL & 7 &  0.59 &  0.61 & 2.1 & & 1978 & TX & 24 & -0.73 & -0.46 & -5.5\\
2014 & SC & 7 &  0.59 &  0.60 & 2.1 & & 1980 & TX & 24 & -0.72 & -0.46 & -5.5\\
1994 & WA & 9 &  0.58 &  0.53 & 2.4 & & 1974 & TX & 24 & -0.71 & -0.45 & -5.4\\
2016 & TX & 36 &  0.58 &  0.32 & 5.8 & & 1972 & TX & 24 & -0.70 & -0.44 & -5.3\\
2012 & NC & 13 &  0.58 &  0.45 & 2.9 & & 1992 & TX & 30 & -0.70 & -0.41 & -6.1\\
2012 & VA & 11 &  0.58 &  0.48 & 2.6 & & 1972 & MO & 10 & -0.69 & -0.60 & -3.0\\
1974 & OH & 23 &  0.57 &  0.37 & 4.2 & & 1992 & WA & 9 & -0.68 & -0.62 & -2.8\\
2012 & AL & 7 &  0.57 &  0.59 & 2.1 & & 2014 & MD & 8 & -0.65 & -0.63 & -2.5\\
2016 & SC & 7 &  0.56 &  0.58 & 2.0 & & 1990 & TX & 27 & -0.63 & -0.38 & -5.2\\
2012 & MI & 14 &  0.56 &  0.43 & 3.0 & & 1978 & WA & 7 & -0.61 & -0.63 & -2.2\\
2010 & FL & 25 &  0.56 &  0.35 & 4.4 & & 1988 & TX & 27 & -0.60 & -0.36 & -4.9\\
2002 & FL & 25 &  0.56 &  0.35 & 4.4 & & 1994 & TX & 30 & -0.59 & -0.35 & -5.2\\
2006 & VA & 11 &  0.55 &  0.46 & 2.5 & & 1976 & WA & 7 & -0.59 & -0.60 & -2.1\\
2016 & OH & 16 &  0.54 &  0.39 & 3.1 & & 1980 & MD & 8 & -0.58 & -0.56 & -2.2\\
1994 & OK & 6 &  0.54 &  0.60 & 1.8 & & 2016 & MD & 8 & -0.58 & -0.56 & -2.2\\
2014 & OH & 16 &  0.53 &  0.38 & 3.1 & & 1980 & MA & 12 & -0.57 & -0.46 & -2.8\\
2006 & MI & 15 &  0.53 &  0.39 & 2.9 & & 2014 & CA & 53 & -0.57 & -0.29 & -7.6\\
2004 & FL & 25 &  0.52 &  0.33 & 4.1 & & 1974 & WA & 7 & -0.56 & -0.58 & -2.0\\
2006 & OH & 18 &  0.52 &  0.36 & 3.3 & & 2012 & MD & 8 & -0.55 & -0.53 & -2.1\\
1998 & AZ & 6 &  0.52 &  0.58 & 1.8 & & 1978 & FL & 15 & -0.54 & -0.40 & -3.0\\
2000 & AZ & 6 &  0.51 &  0.57 & 1.7 & & 1978 & OK & 6 & -0.54 & -0.60 & -1.8\\
2010 & IL & 19 &  0.51 &  0.35 & 3.3 & & 1978 & NC & 11 & -0.54 & -0.45 & -2.5\\
2014 & MI & 14 &  0.50 &  0.38 & 2.7 & & 1980 & OK & 6 & -0.54 & -0.60 & -1.8\\
\hline
\end{tabular}
  \label{tab:ext1972}
\end{table}

\begin{landscape}
\begin{table}[ht]
  \centering
  \caption{Values of declination for congressional elections. Empty
    entries are due to one party winning all of the seats.}
  \resizebox{\textwidth}{!}{%
  \begin{tabular}{@{}lrrrrrrrrrrrrrrrrrrrrrrr@{}} \hline
& 1972 & 1974 & 1976 & 1978 & 1980 & 1982 & 1984 & 1986 & 1988 & 1990 & 1992 & 1994 & 1996 & 1998 & 2000 & 2002 & 2004 & 2006 & 2008 & 2010 & 2012 & 2014 & 2016\\\hline
AK &   &   &   &   &   &   &   &   &   &   &   &   &   &   &   &   &   &   &   &   &   &   &  \\
AL &  0.01 &  0.15 &  0.18 &  0.12 &  0.11 &  0.09 &  0.06 &  -0.10 &  -0.05 &  0.07 &  0.22 &  -0.25 &  0.32 &  0.42 &  0.34 &  0.36 &  0.37 &  0.41 &  -0.02 &  0.61 &  0.59 &  0.51 &  0.54\\
AR &  -0.45 &  0.33 &  -0.23 &  0.06 &  -0.09 &  0.11 &  -0.51 &  -0.38 &  -0.36 &  -0.40 &  0.41 &  -0.10 &  -0.17 &  0.01 &  -0.38 &  -0.42 &  -0.24 &  -0.21 &  -0.27 &  0.19 &   &   &  \\
AZ &  0.39 &  0.42 &  0.05 &  -0.16 &  -0.11 &  0.01 &  0.41 &  0.52 &  0.45 &  0.42 &  -0.16 &  0.48 &  0.52 &  0.58 &  0.57 &  0.36 &  0.43 &  0.05 &  -0.09 &  0.07 &  -0.28 &  -0.35 &  0.01\\
CA &  0.05 &  0.01 &  -0.10 &  -0.06 &  -0.02 &  -0.20 &  -0.18 &  -0.11 &  -0.08 &  -0.01 &  0.08 &  0.00 &  0.02 &  0.14 &  -0.02 &  -0.12 &  -0.06 &  0.00 &  0.13 &  -0.06 &  -0.09 &  -0.28 &  -0.10\\
CO &  0.06 &  0.01 &  -0.33 &  -0.24 &  -0.36 &  -0.10 &  -0.04 &  -0.23 &  -0.01 &  0.05 &  0.28 &  -0.08 &  0.04 &  0.08 &  0.20 &  0.30 &  0.17 &  0.14 &  -0.18 &  0.09 &  0.19 &  0.05 &  0.11\\
CT &  -0.07 &  0.15 &  -0.29 &  -0.34 &  -0.15 &  -0.00 &  -0.14 &  0.21 &  -0.03 &  -0.13 &  -0.12 &  -0.05 &  0.02 &  -0.26 &  0.16 &  0.30 &  0.43 &  0.11 &   &   &   &   &  \\
DE &   &   &   &   &   &   &   &   &   &   &   &   &   &   &   &   &   &   &   &   &   &   &  \\
FL &  -0.35 &  -0.14 &  -0.12 &  -0.40 &  -0.25 &  -0.00 &  -0.28 &  -0.13 &  0.02 &  0.03 &  0.06 &  0.09 &  0.20 &  0.22 &  0.28 &  0.29 &  0.35 &  0.29 &  0.19 &  0.29 &  0.27 &  0.15 &  0.06\\
GA &  -0.66 &   &   &  -0.15 &  -0.38 &  -0.24 &  -0.23 &  -0.04 &  -0.44 &  0.14 &  0.05 &  0.18 &  0.30 &  0.30 &  0.24 &  0.04 &  -0.03 &  -0.05 &  0.09 &  0.07 &  0.21 &  0.31 &  0.36\\
HI &   &   &   &   &   &   &   &  0.27 &  0.38 &   &   &   &   &   &   &   &   &   &   &   &   &   &  \\
IA &  0.07 &  0.02 &  0.08 &  -0.01 &  0.06 &  0.11 &  0.19 &  0.21 &  0.30 &  0.26 &  0.43 &   &  -0.05 &  0.25 &  0.43 &  0.14 &  0.15 &  -0.22 &  -0.11 &  -0.41 &  0.07 &  0.03 &  0.06\\
ID &   &   &   &   &   &   &  -0.40 &  -0.27 &  -0.01 &   &  -0.05 &   &   &   &   &   &   &   &  -0.43 &   &   &   &  \\
IL &  0.22 &  0.20 &  0.12 &  0.13 &  0.21 &  0.15 &  -0.14 &  -0.06 &  -0.19 &  -0.20 &  0.02 &  -0.02 &  0.22 &  0.16 &  0.22 &  0.17 &  0.13 &  0.26 &  0.11 &  0.35 &  -0.11 &  -0.01 &  -0.04\\
IN &  0.04 &  -0.29 &  -0.23 &  -0.11 &  -0.07 &  -0.02 &  -0.10 &  -0.15 &  -0.10 &  -0.36 &  -0.19 &  -0.12 &  0.04 &  0.01 &  0.04 &  0.05 &  0.31 &  -0.10 &  0.01 &  -0.02 &  0.44 &  0.21 &  0.31\\
KS &  0.21 &  0.18 &  -0.31 &  0.51 &  0.50 &  0.12 &  0.09 &  0.08 &  0.08 &  0.23 &  -0.19 &   &   &  -0.16 &  -0.19 &  -0.22 &  -0.02 &  -0.25 &  0.10 &   &   &   &  \\
KY &  -0.47 &  -0.01 &  -0.13 &  -0.01 &  -0.02 &  0.19 &  -0.07 &  0.03 &  -0.22 &  -0.16 &  -0.12 &  -0.12 &  0.23 &  0.09 &  0.18 &  -0.11 &  0.31 &  0.04 &  0.16 &  -0.26 &  0.44 &  0.38 &  0.36\\
LA &  -0.12 &  0.14 &  -0.36 &  -0.13 &  -0.34 &  -0.24 &  -0.35 &  0.14 &  0.11 &  0.04 &  0.04 &  -0.27 &  0.30 &  0.31 &  0.27 &  -0.03 &  0.26 &  0.01 &  0.36 &  0.45 &  0.45 &  0.46 &  0.41\\
MA &  -0.21 &  -0.41 &  -0.34 &  -0.43 &  -0.48 &  -0.63 &  -0.65 &  -0.64 &  -0.66 &  -0.68 &  0.03 &  0.05 &   &   &   &   &   &   &   &   &   &   &  \\
MD &  0.10 &  0.13 &  0.24 &  -0.16 &  -0.56 &  -0.37 &  -0.08 &  0.07 &  -0.27 &  -0.12 &  0.20 &  0.08 &  0.22 &  0.19 &  0.26 &  -0.41 &  -0.34 &  -0.18 &  -0.32 &  -0.12 &  -0.53 &  -0.63 &  -0.56\\
ME &  0.12 &   &   &   &   &   &   &  -0.40 &  -0.06 &  0.22 &  0.24 &  0.02 &   &   &   &   &   &   &   &   &   &  0.29 &  0.08\\
MI &  0.26 &  0.19 &  0.15 &  -0.14 &  -0.08 &  -0.08 &  -0.12 &  0.04 &  -0.11 &  -0.04 &  -0.11 &  -0.18 &  -0.11 &  -0.18 &  0.08 &  0.24 &  0.27 &  0.39 &  0.12 &  0.14 &  0.43 &  0.38 &  0.34\\
MN &  0.16 &  0.09 &  -0.09 &  0.00 &  0.24 &  -0.02 &  -0.11 &  0.15 &  0.07 &  -0.26 &  -0.28 &  -0.44 &  -0.25 &  -0.33 &  0.01 &  0.08 &  0.15 &  0.01 &  0.21 &  0.09 &  0.11 &  -0.12 &  -0.09\\
MO &  -0.60 &  -0.01 &  -0.32 &  -0.17 &  -0.02 &  0.12 &  -0.28 &  0.14 &  0.13 &  0.01 &  -0.09 &  -0.29 &  0.10 &  -0.06 &  0.17 &  0.01 &  -0.02 &  0.12 &  0.15 &  0.07 &  0.40 &  0.25 &  0.33\\
MS &  0.27 &  -0.14 &  -0.10 &  -0.05 &  -0.02 &  -0.05 &  -0.14 &  -0.55 &  0.02 &   &   &  -0.38 &  0.01 &  -0.07 &  -0.17 &  -0.04 &  -0.12 &  0.11 &  -0.24 &  0.33 &  0.36 &  0.25 &  0.33\\
MT &  0.42 &   &  0.24 &  0.01 &  0.05 &  0.15 &  0.02 &  0.19 &  0.12 &  -0.04 &   &   &   &   &   &   &   &   &   &   &   &   &  \\
NC &  -0.16 &  -0.05 &  -0.16 &  -0.45 &  -0.01 &  -0.36 &  0.08 &  0.06 &  -0.32 &  -0.09 &  -0.25 &  0.07 &  -0.13 &  -0.02 &  0.10 &  0.02 &  0.06 &  0.02 &  -0.06 &  -0.25 &  0.45 &  0.54 &  0.48\\
ND &   &   &   &   &   &   &   &   &   &   &   &   &   &   &   &   &   &   &   &   &   &   &  \\
NE &   &   &  -0.23 &  -0.24 &   &   &   &   &  -0.36 &  0.13 &  -0.24 &   &   &   &   &   &   &   &   &   &   &  -0.30 &  \\
NH &   &  -0.28 &  0.16 &  -0.10 &  -0.07 &  -0.31 &   &   &   &  -0.07 &  0.17 &   &   &   &   &   &   &   &   &   &   &  0.08 &  \\
NJ &  -0.03 &  -0.18 &  -0.27 &  -0.11 &  -0.04 &  0.00 &  -0.08 &  -0.05 &  -0.15 &  -0.19 &  -0.10 &  0.20 &  0.15 &  0.05 &  0.10 &  0.05 &  0.09 &  0.22 &  0.06 &  -0.01 &  0.28 &  0.17 &  0.07\\
NM &  0.32 &  0.16 &  -0.03 &  0.10 &   &  0.36 &  0.05 &  0.28 &  0.44 &  0.30 &  0.37 &  0.12 &  0.38 &  0.08 &  0.39 &  0.37 &  0.40 &  0.53 &   &  -0.11 &  -0.11 &  -0.27 &  -0.17\\
NV &   &   &   &   &   &  0.00 &  -0.31 &  -0.08 &  0.14 &  0.01 &  0.20 &   &   &  -0.27 &  -0.31 &  -0.13 &  0.27 &  0.48 &  0.01 &  0.28 &  0.06 &  0.24 &  -0.35\\
NY &  -0.00 &  -0.05 &  -0.15 &  -0.13 &  -0.02 &  0.08 &  -0.00 &  0.03 &  -0.09 &  -0.04 &  0.07 &  0.02 &  0.14 &  0.06 &  -0.01 &  -0.15 &  -0.01 &  0.08 &  -0.44 &  -0.00 &  0.09 &  -0.03 &  0.14\\
OH &  0.39 &  0.37 &  0.12 &  0.11 &  0.02 &  0.20 &  -0.11 &  -0.04 &  -0.02 &  0.04 &  -0.04 &  0.22 &  0.08 &  0.14 &  0.14 &  0.23 &  0.31 &  0.36 &  0.01 &  0.28 &  0.55 &  0.38 &  0.39\\
OK &  -0.53 &   &  0.16 &  -0.61 &  -0.60 &  -0.46 &  -0.59 &  -0.02 &  -0.04 &  -0.02 &  0.19 &  0.60 &   &   &  0.21 &  0.55 &  0.39 &  0.50 &  0.43 &  0.08 &   &   &  \\
OR &  0.00 &   &   &   &  0.20 &  0.17 &  0.05 &  0.03 &  0.22 &  -0.43 &  -0.41 &  0.12 &  -0.39 &  -0.39 &  -0.52 &  -0.53 &  -0.55 &  -0.46 &  -0.44 &  -0.60 &  -0.48 &  -0.51 &  -0.53\\
PA &  -0.08 &  0.15 &  -0.15 &  -0.18 &  -0.03 &  -0.03 &  -0.09 &  0.01 &  -0.03 &  0.06 &  -0.00 &  -0.07 &  0.06 &  -0.01 &  0.14 &  0.24 &  0.28 &  0.10 &  -0.05 &  0.21 &  0.53 &  0.43 &  0.47\\
RI &   &   &   &   &  0.26 &  0.13 &  0.01 &  -0.27 &   &  0.09 &  0.00 &   &   &   &   &   &   &   &   &   &   &   &  \\
SC &  0.01 &  -0.24 &  -0.27 &  0.10 &  0.22 &  0.05 &  -0.01 &  0.19 &  -0.06 &  -0.14 &  0.07 &  -0.00 &  0.13 &  0.27 &  0.24 &  0.12 &  0.17 &  0.13 &  0.31 &  0.46 &  0.65 &  0.60 &  0.58\\
SD &  0.13 &   &   &  -0.15 &  0.15 &   &   &   &   &   &   &   &   &   &   &   &   &   &   &   &   &   &  \\
TN &  0.27 &  -0.11 &  -0.02 &  -0.23 &  -0.18 &  -0.10 &  -0.34 &  -0.27 &  -0.29 &  -0.25 &  -0.30 &  -0.08 &  0.10 &  0.11 &  0.09 &  -0.18 &  -0.13 &  -0.02 &  -0.13 &  0.29 &  0.38 &  0.34 &  0.34\\
TX &  -0.44 &  -0.45 &  -0.67 &  -0.46 &  -0.46 &  -0.47 &  -0.20 &  -0.14 &  -0.37 &  -0.39 &  -0.41 &  -0.35 &  -0.17 &  -0.19 &  -0.16 &  -0.20 &  0.22 &  0.18 &  0.23 &  0.20 &  0.24 &  0.24 &  0.32\\
UT &   &   &  0.06 &  -0.26 &   &   &   &  0.04 &  0.04 &  0.03 &  -0.43 &  0.14 &   &   &  -0.02 &  -0.28 &  -0.10 &  0.09 &  0.12 &  -0.32 &  -0.33 &   &  \\
VA &  0.35 &  0.19 &  0.23 &  -0.04 &  0.58 &  -0.02 &  -0.04 &  0.04 &  -0.08 &  -0.08 &  -0.17 &  -0.21 &  -0.05 &  -0.08 &  0.25 &  0.25 &  0.29 &  0.46 &  -0.03 &  0.27 &  0.48 &  0.32 &  0.21\\
VT &   &   &   &   &   &   &   &   &   &  &  &  &  &  &  &  &  &   &   &   &   &   &  \\
WA &  0.22 &  -0.58 &  -0.60 &  -0.63 &  -0.45 &  -0.16 &  -0.10 &  0.02 &  -0.16 &  -0.06 &  -0.62 &  0.53 &  0.42 &  -0.03 &  -0.16 &  -0.23 &  0.02 &  0.11 &  -0.03 &  -0.04 &  -0.07 &  -0.07 &  -0.04\\
WI &  0.16 &  -0.15 &  -0.33 &  -0.19 &  -0.05 &  -0.01 &  -0.22 &  -0.12 &  -0.16 &  0.16 &  0.10 &  0.00 &  -0.18 &  -0.02 &  -0.18 &  -0.07 &  -0.09 &  -0.15 &  -0.14 &  0.07 &  0.31 &  0.19 &  0.20\\
WV &   &   &   &   &  0.32 &   &   &   &   &   &   &   &   &   &  0.28 &  -0.04 &  0.00 &  0.04 &  0.06 &  0.03 &  -0.14 &   &  \\
WY &   &   &   &   &   &   &   &   &   &   &   &   &   &   &   &   &   &   &   &   &   &   &  \\
\hline
\end{tabular}}\label{tab:cong}
\end{table}

\newpage

\begin{table}[ht]
  \centering
  \caption{Values of declination for state lower house
    elections. Empty entries are due to one party winning all of the
    seats or to the election containing multi-member districts.}
  \resizebox{\textwidth}{!}{%
  \begin{tabular}{@{}lrrrrrrrrrrrrrrrrrrrrrrr@{}} \hline
& 1972 & 1974 & 1976 & 1978 & 1980 & 1982 & 1984 & 1986 & 1988 & 1990 & 1992 & 1994 & 1996 & 1998 & 2000 & 2002 & 2004 & 2006 & 2008 & 2010\\\midrule
AK &  &  &  &  &  &  &  &  &  &  &  0.00 &  0.01 &  -0.02 &  0.03 &  0.13 &  0.06 &  0.09 &  0.05 &  -0.06 &  -0.16\\
AL &  &   &  &  -0.61 &  &  &  &  -0.50 &  &  -0.40 &  &  -0.33 &  &  -0.15 &  &  -0.15 &  &  -0.06 &  &  0.01\\
AR &  &  &  &  &  &  &  &  &  &  &  &  &  &  &  &  -0.18 &  -0.25 &  -0.20 &  -0.16 &  -0.10\\
AZ &  &  &  &  &  &  &  &  &  &  &  &  &  &  &  &  &  &  &  & \\
CA &  -0.11 &  -0.06 &  -0.13 &  -0.06 &  -0.11 &  -0.04 &  -0.14 &  -0.00 &  -0.09 &  0.00 &  0.02 &  0.08 &  0.09 &  0.05 &  -0.01 &  -0.04 &  -0.02 &  0.03 &  0.12 &  -0.07\\
CO &  0.06 &  0.11 &  0.11 &  0.13 &  0.07 &  0.13 &  0.30 &  0.13 &  0.11 &  0.16 &  0.03 &  0.06 &  0.11 &  0.13 &  0.11 &  0.06 &  -0.04 &  -0.04 &  -0.02 &  -0.05\\
CT &  0.13 &  -0.14 &  -0.07 &  -0.08 &  -0.04 &  0.01 &  -0.00 &  -0.01 &  -0.04 &  -0.09 &  0.17 &  -0.02 &  -0.04 &  -0.08 &  -0.05 &  -0.02 &  -0.07 &  -0.03 &  -0.07 &  -0.10\\
DE &  0.04 &  -0.02 &  -0.11 &  0.02 &  0.10 &  -0.23 &  0.01 &  -0.01 &  -0.06 &  0.08 &  0.06 &  0.17 &  0.19 &  0.22 &  0.29 &  0.40 &  0.29 &  0.27 &  -0.00 &  -0.14\\
FL &  &  &  &  &  &  -0.16 &  -0.20 &  -0.12 &  -0.14 &  -0.08 &  -0.07 &  -0.10 &  0.03 &  0.16 &  0.25 &  0.26 &  0.34 &  0.27 &  0.26 &  0.21\\
GA &  &  &  &  &  &  &  &  &  &  &  -0.19 &  -0.18 &  -0.06 &  -0.02 &  -0.04 &  &  -0.06 &  0.07 &  0.05 &  -0.05\\
HI &  &  &  &  &  &  -0.32 &  -0.22 &  -0.25 &  -0.43 &  -0.29 &  -0.66 &  -0.54 &  -0.30 &  -0.25 &  0.01 &  -0.18 &  -0.37 &  -0.31 &  -0.46 &  -0.38\\
IA &  0.05 &  -0.02 &  -0.06 &  0.02 &  0.05 &  -0.06 &  -0.12 &  -0.04 &  -0.08 &  -0.01 &  0.06 &  0.10 &  0.04 &  0.11 &  0.13 &  0.04 &  0.02 &  0.05 &  -0.05 &  -0.02\\
ID &  &  &  &  &  &  &  &  &  &  &  &  &  &  &  &  &  &  &  & \\
IL &  &  &  &  &  &  -0.01 &  -0.04 &  0.05 &  -0.00 &  -0.02 &  0.08 &  0.15 &  0.19 &  0.15 &  0.14 &  0.06 &  0.11 &  0.17 &  0.11 &  0.00\\
IN &  &  &  &  &  &  &  &  &  &  &  -0.10 &  -0.06 &  -0.04 &  -0.07 &  -0.07 &  -0.12 &  -0.04 &  0.06 &  -0.01 &  0.01\\
KS &  0.13 &  0.17 &  0.04 &  0.06 &  -0.00 &  0.09 &  0.07 &  0.10 &  -0.00 &  -0.02 &  -0.04 &  -0.01 &  0.00 &  0.09 &  0.18 &  0.10 &  0.14 &  0.20 &  0.11 &  0.14\\
KY &  &  &  &  &  &  &  -0.30 &  -0.20 &  -0.26 &  -0.17 &  -0.19 &  -0.15 &  -0.11 &  -0.10 &  -0.11 &  -0.13 &  -0.06 &  0.02 &  -0.09 &  -0.14\\
LA &  -0.36 &  &  &  &  &  &  &  &  &  &  &  &  &  &  &  &  &  &  & \\
MA &  &  -0.19 &  -0.26 &  -0.26 &  -0.28 &  -0.34 &  -0.32 &  -0.29 &  -0.33 &  -0.23 &  -0.17 &  -0.27 &  -0.29 &  -0.24 &  -0.33 &  -0.37 &  -0.41 &  -0.37 &  -0.50 &  -0.20\\
MD &  &  &  &  &  &  &  &  &  &  &  &  &  &  &  &  &  &  &  & \\
ME &  &  &  &  0.02 &  -0.04 &  -0.03 &  -0.05 &  -0.08 &  -0.16 &  -0.06 &  -0.02 &  0.00 &  0.04 &  0.04 &  -0.07 &  0.01 &  0.10 &  0.03 &  -0.03 &  -0.02\\
MI &  -0.02 &  0.15 &  0.01 &  -0.05 &  -0.03 &  0.11 &  -0.03 &  0.03 &  -0.01 &  0.05 &  0.15 &  0.04 &  0.10 &  0.17 &  0.22 &  0.25 &  0.20 &  0.21 &  0.09 &  0.19\\
MN &  &  -0.20 &  -0.19 &  0.04 &  0.01 &  0.02 &  0.09 &  -0.07 &  -0.06 &  -0.02 &  -0.05 &  -0.06 &  0.06 &  0.07 &  0.11 &  0.25 &  0.15 &  0.02 &  -0.01 &  0.13\\
MO &  &  0.02 &  -0.02 &  -0.11 &  -0.13 &  -0.10 &  -0.19 &  -0.18 &  &  0.04 &  -0.04 &  -0.04 &  0.02 &  0.01 &  0.05 &  0.06 &  0.14 &  0.17 &  0.09 &  0.09\\
MS &  &  &  &  &  &  &  &  &  &  &  &  &  &  &  &  &  &  &  & \\
MT &  &  -0.08 &  -0.02 &  -0.03 &  0.05 &  -0.03 &  -0.04 &  0.04 &  0.02 &  -0.03 &  0.06 &  0.09 &  0.13 &  0.09 &  0.02 &  0.03 &  -0.04 &  0.02 &  0.01 &  0.07\\
NC &  &  &  &  &  &  &  &  &  &  &  &  &  &  &  &  0.04 &  -0.00 &  0.02 &  0.01 &  -0.00\\
ND &  &  &  &  &  &  &  &  &  &  &  &  &  &  &  &  &  &  &  & \\
NE &  &  &  &  &  &  &  &  &  &  &  &  &  &  &  &  &  &  &  & \\
NH &  &  &  &  &  &  &  &  &  &  &  &  &  &  &  &  &  &  &  & \\
NJ &  &  &  &  &  &  &  &  &  &  &  &  &  &  &  &  &  &  &  & \\
NM &  -0.32 &  -0.16 &  -0.17 &  -0.02 &  -0.15 &  -0.06 &  -0.08 &  -0.16 &  -0.08 &  -0.20 &  -0.23 &  -0.17 &  -0.03 &  -0.01 &  -0.08 &  -0.09 &  0.02 &  0.10 &  -0.01 &  0.05\\
NV &  0.09 &  -0.24 &  -0.47 &  0.05 &  -0.13 &  0.03 &  0.15 &  -0.25 &  -0.32 &  -0.00 &  -0.26 &  0.01 &  -0.12 &  -0.24 &  -0.21 &  0.01 &  -0.07 &  -0.01 &  -0.06 &  -0.08\\
NY &  0.20 &  0.15 &  0.10 &  0.08 &  0.04 &  -0.00 &  -0.06 &  0.05 &  -0.01 &  0.07 &  -0.05 &  -0.06 &  0.08 &  0.04 &  0.07 &  -0.01 &  0.02 &  0.09 &  0.05 &  0.00\\
OH &  -0.04 &  0.07 &  -0.02 &  -0.08 &  -0.06 &  0.01 &  -0.12 &  -0.07 &  -0.10 &  -0.05 &  0.06 &  0.08 &  0.24 &  0.24 &  0.23 &  0.25 &  0.25 &  0.27 &  0.09 &  0.14\\
OK &  -0.19 &  -0.21 &  -0.32 &  -0.23 &  -0.24 &  -0.27 &  -0.24 &  -0.18 &  -0.19 &  -0.11 &  -0.22 &  -0.24 &  -0.21 &  -0.10 &  0.05 &  -0.08 &  0.06 &  0.14 &  0.16 &  0.16\\
OR &  0.02 &  -0.03 &  -0.03 &  -0.04 &  -0.02 &  -0.10 &  -0.10 &  0.01 &  -0.01 &  0.08 &  0.11 &  0.03 &  0.08 &  0.17 &  0.13 &  0.16 &  0.16 &  0.11 &  0.02 &  0.01\\
PA &  0.02 &  0.10 &  -0.00 &  0.05 &  0.02 &  0.09 &  0.02 &  0.05 &  0.02 &  0.04 &  0.06 &  -0.03 &  0.08 &  0.07 &  0.13 &  0.05 &  0.13 &  0.12 &  0.06 &  0.03\\
RI &  -0.10 &  -0.10 &  -0.22 &  -0.27 &  -0.18 &  -0.32 &  -0.23 &  -0.26 &  -0.28 &  -0.38 &  -0.16 &  -0.38 &  -0.25 &  -0.37 &  -0.23 &  -0.36 &  -0.16 &  -0.04 &  -0.35 &  -0.33\\
SC &  &  -0.26 &  -0.46 &  -0.39 &  -0.43 &  -0.36 &  -0.40 &  -0.19 &  -0.18 &  -0.09 &  -0.14 &  -0.09 &  0.09 &  0.06 &  0.08 &  0.13 &  0.08 &  0.13 &  0.11 &  0.05\\
SD &  &  &  &  &  &  &  &  &  &  &  &  &  &  &  &  &  &  &  & \\
TN &  -0.01 &  -0.01 &  -0.08 &  -0.02 &  -0.08 &  -0.06 &  -0.17 &  -0.04 &  -0.04 &  0.00 &  -0.18 &  -0.25 &  -0.16 &  -0.12 &  -0.10 &  -0.10 &  -0.08 &  0.01 &  0.01 &  0.11\\
TX &  &  &  -0.53 &  -0.52 &  -0.31 &  -0.30 &  -0.22 &  -0.11 &  -0.10 &  -0.11 &  -0.15 &  -0.24 &  -0.07 &  -0.03 &  -0.01 &  0.07 &  0.02 &  0.05 &  -0.06 &  0.05\\
UT &  0.01 &  0.04 &  -0.05 &  0.10 &  0.12 &  0.16 &  0.21 &  0.11 &  0.01 &  0.03 &  0.05 &  0.06 &  0.15 &  0.15 &  0.08 &  0.12 &  0.16 &  0.20 &  0.11 &  0.15\\
VA &  &  &  &  &  &  &  &  &  &  &  &  &  &  &  &  &  &  &  & \\
VT &  &  &  &  &  &  &  &  &  &  &  &  &  &  &  &  &  &  &  & \\
WA &  &  &  &  &  &  &  &  &  &  &  &  &  &  &  &  &  &  &  & \\
WI &  -0.11 &  -0.07 &  -0.08 &  -0.09 &  -0.08 &  0.05 &  -0.02 &  -0.04 &  -0.02 &  -0.07 &  0.00 &  -0.05 &  0.01 &  0.10 &  0.13 &  0.10 &  0.17 &  0.14 &  0.08 &  0.03\\
WV &  &  &  &  &  &  &  &  &  &  &  &  &  &  &  &  &  &  &  & \\
WY &  &  &  &  &  &  &  &  &  &  &  0.18 &  0.11 &  0.05 &  0.14 &  0.23 &  0.14 &  0.18 &  0.18 &  0.06 &  0.26\\
\hline
\end{tabular}}\label{tab:state}
\end{table}

\end{landscape}




\bibliography{gerrymandering}

\bibliographystyle{alpha}

\end{document}